\documentclass[twocolumn,aps,showpacs,floatfix,superscriptaddress]{revtex4}

\usepackage{amsmath, amssymb, amsfonts}
\usepackage{bm, mathrsfs, dsfont,bbold}
\usepackage{braket}
\usepackage{graphicx}
\usepackage{multirow}
\usepackage{color}
\usepackage{array}

\def\ie{{\it i.e.},\ }
\def\eg{{\it e.g.}\ }

\newcommand{\up}{\uparrow}
\newcommand{\dn}{\downarrow}

\newcommand{\Tr}{\mathrm{Tr}}

\graphicspath{{./}{./pics/}{../pics/Sec4/}{./pics/Raw/}{./pics/make/}}

\begin{document}

\title{Coherent control of an NV$^{-}$ center with one adjacent  $^{13}$C}

\author{Burkhard Scharfenberger}\email{burkhard@nii.ac.jp}
\affiliation{National Institute of Informatics, 2-1-2 Hitotsubashi, Chiyoda-ku, Tokyo 101-8430, Japan}

\author{William J. Munro}
\affiliation{NTT Basic Research Laboratories, NTT Corporation,\\ 3-1 Morinosato Wakamiya, Atsugi, Kanagawa 243-0198, Japan}
\affiliation{National Institute of Informatics, 2-1-2 Hitotsubashi, Chiyoda-ku, Tokyo 101-8430, Japan}

\author{Kae Nemoto}
\affiliation{National Institute of Informatics, 2-1-2 Hitotsubashi, Chiyoda-ku, Tokyo 101-8430, Japan}

\date{\today}
\pagestyle{plain}

\begin{abstract}
 We investigate the theoretically achievable fidelities when coherently controlling an effective 
 three qubit system consisting of a negatively charged nitrogen vacancy (NV$^-$) center in 
 diamond with an additional nearby carbon $^{13}$C spin $I_{\text{C}}=1/2$ via square radio 
 and microwave frequency pulses in different magnetic field regimes. 
 Such a system has potentially interesting applications in quantum information related tasks 
 such as distributed quantum computation or quantum repeater schemes.
 We find that the best fidelities can be achieved in an intermediate magnetic field regime.
 However,  with only square pulses it will be challenging to reach the fidelity 
 threshold(s) predicted by current models of fault-tolerant quantum computing.
\end{abstract}


\maketitle

\section{Introduction}

Impurity spins in solids have long been known for their potential to be used in quantum 
information processing devices~\cite{weber-10pnas8513,Kane98n393,Ohlsson-02oc71}. 
Among these, the negatively charged vacancy centers(NV$^-$) in diamond has stood 
out for its exceptional properties. 
It is a well localized, stable and optically controllable spin in the 'vacuum' of a mostly 
spin-less carbon lattice~\cite{Doherty-13physrep1}. Given these virtues, they have early 
been recognized as a good solid state qubit, showing long coherence times even at room
temperature~\cite{davies,Kennedy-03apl4190, Jelezko-04prl076401,Bala-09nm383}. 
Moreover, NV$^-$ centers have been employed in a host of applications beyond quantum 
information, ranging from use as single photon source~\cite{kurtsiefer-00prl290, brouri-00ol1294,beveratos-01pra061802}, 
high-resolution sensor in electrometry~\cite{dolde-11np459}, magnetrometry 
~\cite{degen08apl243111,bala-08n648,maze-08n644,taylor-08np810,Bala-09nm383,
Maertz-10apl092504,Hall-10prb045208,Cooper-14nc3141}, 
decoherence microscopy~\cite{Cole-09nano495401,Hall-09prl220802,Hall-10pnas18777}, 
nano-scale NMR sensor~\cite{Zhao-12nn657,Staudacher-13s1231675,Mamin-13s1231540} and 
thermometer~\cite{Kucsko-13n54}.

 % description of NV
A nitrogen vacancy center consists of a vacancy site in a diamond lattice adjacent to a substitutional 
nitrogen atom resulting in a defect of C$_{3,\text{V}}$ symmetry\cite{smith-59pr115.1546,loubser-77diar11}.
In the negative charge state NV$^-$, the electronic wave function is a 
spin $S\!=\!1$ for both a ground state manifold (GSM) with orbital symmetry A$_2$ as well as an excited
state manifold (ESM) of  E-type orbital symmetry separated from the GSM by an optical 637nm (ZPL) transition. 
The NV$^-$ center exhibits the useful properties of optical polarizability and spin dependent fluorescence, 
allowing initialization and readout of the electronic spin even at room temperature. 
These are possible due to the presence of energetically intermediate levels between
the GSM and ESM, which allow spin non-conserving, non-radiative transitions which preferentially
(but not completely) populate the $m_S=0$ sub-level (for a detailed review 
see~\cite{Doherty-13physrep1}). 

% QC applications
Together with the electronic spin of the vacancy, hyperfine-coupled nitrogen and possibly carbon
nuclear spins found in the vicinity can form a quantum register of several qubits.
In such a register, the nuclear spins with their excellent coherence times~\cite{Jelezko-04prl130501,Maurer-12s1283} would serve as quantum memories accessed via the more directly controllable electronic spin of the vacancy.
This system was proposed as node in a quantum repeater~\cite{Childress-05pra052330,Childress-06prl070504}
as well as for quantum information processing~\cite{Yao-2012nc800} and has  
been intensely studied by numerous experiments both at room and at low temperature 
($\approx$4-8K).
The important milestones demonstrated are initialization and single-shot 
readout of electronic and nuclear spins in both temperature regimes~\cite{Dutt-07s1312,Neumann-10s542544,Robledo-11n574},
as well as, at low temperature, creation of entanglement between vacancy electron and nuclear 
spins~\cite{Neumann-08s1326}, the polarization of single photons~\cite{Togan-10n09256} and 
other (distant) NV centers~\cite{Bernien-13n12016}.
Further important steps on the way to a scalable quantum computation architecture are 
a demonstration of room temperature quantum registers formed by long-range dipolar 
coupled NV$^-$ centers~\cite{Neumann-10np249} and entanglement swapping to 
nuclear spins~\cite{Dolde-12n139}.
Moreover, in quantum registers made up of a single NV$^-$ and multiple proximate carbon 
nuclear spins, decoherence-protected operations were performed~\cite{vanderSar-12n82}, 
and recently the first implementations of quantum error correction in diamond-based 
qubits was also demonstrated~\cite{Taminiau-13nn171,Waldherr-13n204}.

%our aim
While these experiments serve as beautiful proofs-of-principle and fidelities achieved are remarkable 
given the practical technical difficulties, they are not yet at thresholds required for scaleable, 
fault-tolerant quantum computation~\cite{Stephens14pra022321}. 
In particular, even with error correction a general computation will require many 
gate executions before the system is reset/corrected and this quickly degrades fidelity.
From the perspective of architecture selection and design, it would be highly desirable to have 
a better theoretical understanding of the ultimate limits to the achievable fidelities, given the
inherent properties of the NV$^-$ system. Previous studies looking at a bare NV$^-$ center in a 
pure carbon lattice have shown that in principle such a system might indeed allow operations with 
high enough accuracy for large-scale quantum computation 
even when using only simple control pulses~\cite{Everitt-13arx1309.3107}, 
at least as long as exciting the vacancy spin out of the GSM is avoided. 
The hyperfine interaction strength in the ESM ($\sim 60$MHz) is relatively stronger than
in the GSM ($\sim 3$MHz) \cite{gali-08prb155206,gali09arx0905.1169v1}., and hence 
any excitation from the GSM could result in dephasing on the nitrogen nuclear spin.  
As quantum information requires not only gate operation but 
also readout and initialization, this difference in coupling strength adds significant constraints 
on the operational regimes of physical parameters and setups.  By contrast, nearby, strongly 
coupled $^{13}$C nuclear spins do not show this difference in hyperfine coupling strength, and 
it might thus be used to design a device immune to this source of dephasing.

This leads to the question investigated in the present work: 
whether high-fidelity control by simple means is still possible
in an effective three-qubit system ($^{15}$NV$^-$+$^{13}$C), where the carbon 
introduces interactions which potentially make high-fidelity control more difficult.

%structure
This paper is structured as follows: in section II we introduce the effective spin model we use
and discuss the magnetic field regimes we investigate it in, which are low magnetic field (low-B)
and intermediate magnetic field (med-B). Of these, we first look at the low-B case in section III, 
investigating single-pulse singe qubit control and entanglement creation via concatenated 
pulses. In section IV we move on to the intermediate magnetic field regime, where multi-qubit
operations can also be achieved with single driving pulses. Section V contains an analysis 
of times and fidelities for derived gates based on the results from the previous section and 
finally we give a concluding discussion in section VI.

\section{Effective spin model}
 The system we study consists of effectively three qubits: the electronic spins of the vacancy defect (V)
 and two nuclear spins, one belonging to the, always present, nitrogen and the other to a nearby 
 carbon $^{13}$C.
 Throughout we will assume the nitrogen to be a $^{15}$N isotope, and thus both nuclei in our system
 have spin $I\!= \! 1/2$, while the electronic spin state is a triplet $S\!=\! 1$.  
 Since we do not consider excitations out of the $^3$A$_2$ GSM, the free time evolution of
 the system is well described by the Hamiltonian~\cite{doherty-12prb205203}:
\begin{equation}
   \begin{aligned}
     H_{\text{NVC}} = &\;H_{\text{V}} + H_{\text{N}} + H_{\text{C}} + H_{\text{VN}} + H_{\text{VC}}  \\
     H_\text{V}    \;\;    = &  \;D S_z^2 + \frac{1}{2} E\left( S_x^2 - S_y^2 \right) + \gamma_\text{e} B S_z   \\
     H_{\text{C/N}}  = &  \; \gamma_\text{C/N} B I_{\text{C/N},z} \\
     H_\text{VN}= & \;  \vec{S}\underbar{A}\vec{I}_\text{N}  \\
     H_\text{VC}  =& \; \vec{S}\underbar{C}\vec{I}_\text{C}\;,
   \end{aligned}
   \label{eq:HNVC}
 \end{equation}
 where $\vec{S}= (S_x, S_y, S_z)^T$ is the vacancy and $\vec{I} = (I_x, I_y, I_z)^T$ the nuclear spin 
 operator and we define the 
 magnetic moments $\gamma_{\text{e}} = g_{\text{e}} \mu_{\text{B}}=28$MHz/mT for the electronic 
 spin as well as the nuclear spins of carbon $\gamma_{\text{C}} = g_{\text{C}} \mu_{\text{n}}=+10.6$kHz/mT 
 and nitrogen $\gamma_{\text{N}} = g_{\text{N}} \mu_{\text{n}}=-4.3$kHz/mT.
 $D$ is a zero-field splitting of 2.88GHz (at low temperature) coming from the spin-spin interaction, 
 $B$ denotes the magnetic field which we assume to be parallel to the NV-axis, and
 $E$ is the crystal strain which is very weak in the GSM ($0...10$MHz) and could be canceled entirely 
 by applying an appropriate electric field. Finally, $\underline{A}$ and $\underline{C}$ are the hyperfine 
 tensors of nitrogen and carbon respectively. 

% hyperfine interactions 
 For symmetry reasons $\underline{A}$ is exactly axial, while $\underline{C}$ is approximately so, 
 even for nearest neighbor carbons where one might expect the contact term to give a significant 
 non-axial contribution. 
 As we consider the nitrogen to be an $^{15}$N isotope ($I\! = \! 1/2$), we do not need to include a 
 nuclear quadrupolar term in~\eqref{eq:HNVC}. 
 Also, the direct dipolar interaction between the two nuclear spins is negligible.

The hyperfine interaction term for the nitrogen consists of parallel and exchange contribution and reads
 $\vec{S}\underbar{A}\vec{I}_\text{N} = A_\parallel S_z I_{\text{N},z} + \frac{1}{2} A_\perp \left(S^+I_{\text{N}}^- +  S^-I_{\text{N}}^+\right)$.
 While the carbon hyperfine-term looks the same in its principal axis system, there are additional
 terms after transforming into NV-adapted coordinates (with z along the NV's symmetry axis):
 \begin{equation}
   \begin{aligned}
       \vec{S}\underbar{C}\vec{I}_\text{C}   \,= &\,\; C_\parallel(\theta)\, S_zI_{\text{C},z} + \frac{1}{2} C_\perp(\theta)\,\left(S^+ I^-_\text{C} + S^- I^+_\text{C} \right) \\
              & \, + \frac{1}{2} C_\text{R}(\theta)\,\left(S^+ I^+_\text{C} + S^- I^-_\text{C} \right) \\
              & \, + C_\Delta(\theta) \left( S_z I_{\text{C},y} + S_y I_{\text{C},z} \right)\; ,
   \end{aligned}
   \label{eq:carbonhf}
\end{equation}
where the C$_\Delta$-term contains z- and y-operators because we used an x-axis rotation in the coordinate
transformation. The four coefficients depend on the angle $\theta$ between the NV axis and the carbon
vacancy axis
% $\vec{r}_\text{VC} = \vec{r}_\text{C}-\vec{r}_\text{V}$ 
and are given by
\begin{equation}
  \begin{aligned}
     C_{||}(\theta) & =  C_{||} \cos^2\theta + C_\perp \sin^2\theta \\
     C_\perp(\theta) & = \frac{1}{2} \left( C_\perp(1 + \cos^2\theta) + C_{||} \sin^2\theta  \right) \\
     C_\text{R}(\theta) & = \frac{1}{2} \left( C_\perp(1 - \cos^2\theta) - C_{||} \sin^2\theta \right) \\  
     C_\Delta(\theta) & = (C_\perp - C_{||}) \sin\theta \cos\theta \,.
  \end{aligned}
  \label{eq:cthetas}
\end{equation}
The effect of the two additional terms $C_\text{R}$ and $C_\Delta$ on energy levels and states
in the magnetic field regime are minimal except that for the $m_S=0$ states at low field,
where $C_\text{R}$ causes a splitting between even parity states ($\ket{0,\up,\up}_\text{VCN}$ and $\ket{0,\dn,\dn}_\text{VCN}$)
while the odd parity states  ($\ket{0,\up,\dn}_\text{VCN}$ and $\ket{0,\dn,\up}_\text{VCN}$) are split
by the exchange term.

The value for $C_\parallel(\theta)$ can be observed directly  in ODMR experiments as the hyperfine 
splitting between different carbon spin orientations. The other parameters are, however, harder to 
confirm.
A rough estimate can be gained by setting the magnetic field to $B=B_\text{x}=103$mT and 
observing the splitting at the avoided crossing between the $m_S=-1$ and $0$ levels. 
Since the Hamiltonian is highly connected, this will not yield good results even for $C_\perp$. 
A better strategy is measuring the level splitting while sweeping the magnetic field
and fitting the model parameters to the obtained data. As an analytic approximation to this, 
one can look at the curvature of the $m_S=0,-1$ levels in a field region around 60-80 mT.
There, at least in 2nd-order perturbation theory, the curvatures are directly proportional to $C^2_\perp$ 
(mixing $\ket{-1,\up}_\text{VC}\leftrightarrow \ket{0,\dn}_\text{VC}$) and $C^2_\text{R}+C^2_\Delta$ 
respectively (mixing $\ket{-1,\dn}_\text{VC}\leftrightarrow \ket{0,\up}_\text{VC}$).

\begin{figure}
  \includegraphics[width=.21\textwidth]{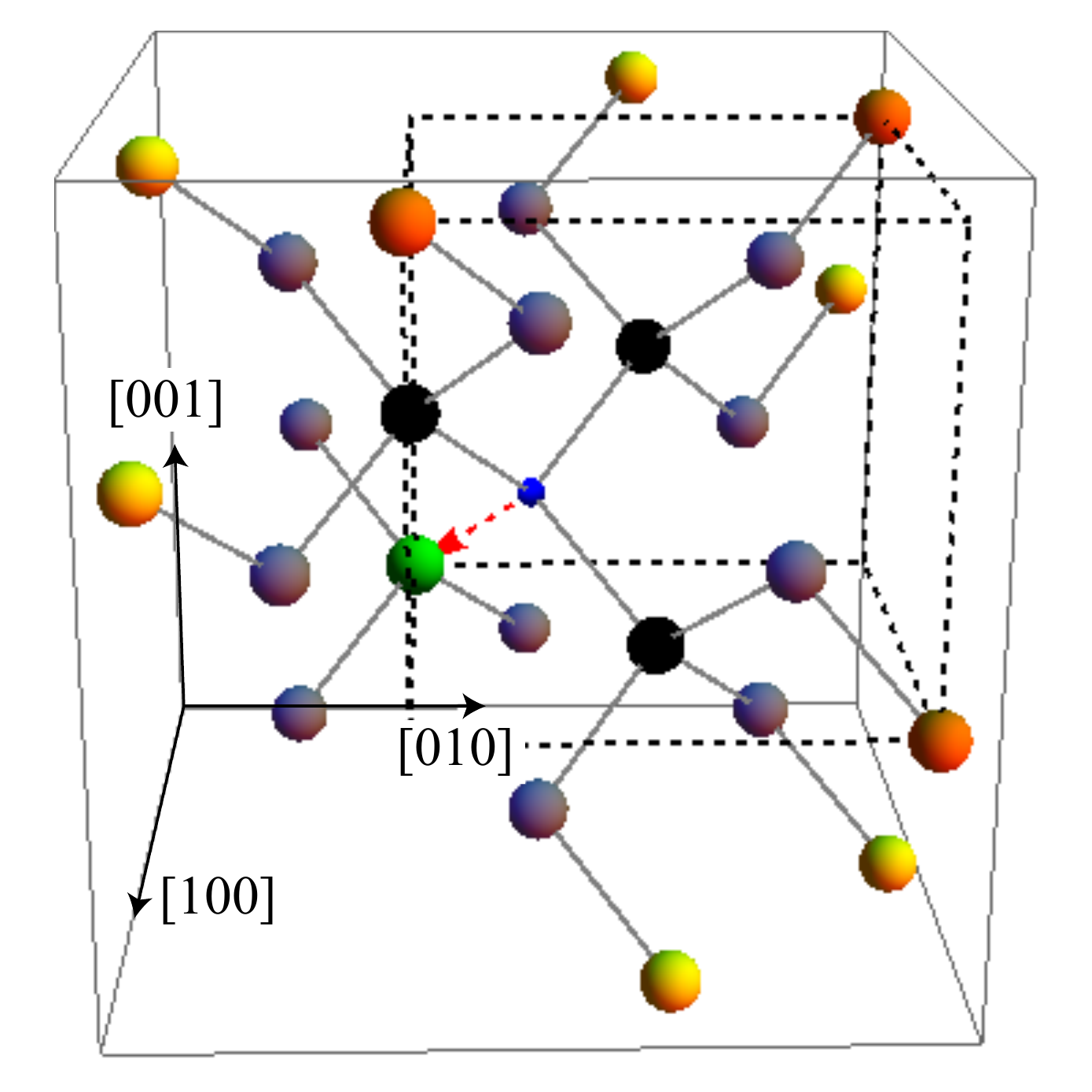} \hspace{.01\textwidth} \includegraphics[width=.215\textwidth]{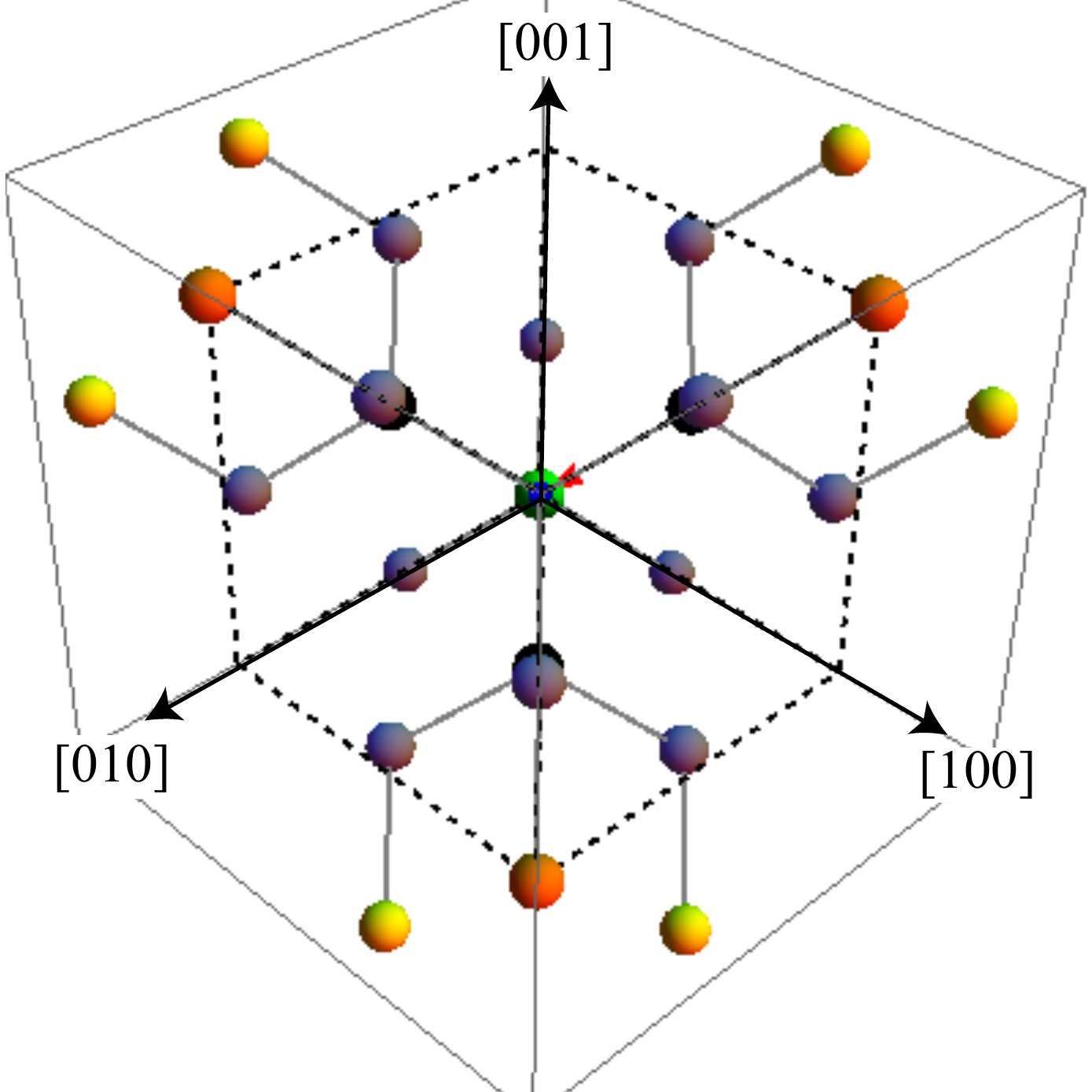}
  \caption{\textbf{Carbon nuclear spin positions}. NV center and the sites where the lattice positions
                  for the (one) carbon $^{13}$C we considered in this study: 
                  on the left a free 3D view and on the right along the [111] direction.
                  The color coding of the spheres is as follows: (small) blue = vacancy, green = nitrogen,
                  black= nearest neighbors (of V), gray = next nearest neighbors, yellow and orange: third 
                  neighbors for which numerical ab-initio calculations suggest strong hyperfine interaction
                  with the vacancy spin due to finite spin density. These calculations find slightly different
                  coupling strength for the two positions yellow and orange, but this has not yet been resolved
                  experimentally.
                  The dashed cage shows a diamond lattice unit-cell.
  		}
   \label{fig:diamond}
\end{figure}

% Carbon positions
We considered two different carbon positions, nearest neighbor and third-neighbor, because these
show the strongest hyperfine interaction and thus offer the potentially fastest gate times. 
For a nearest neighbor carbon, hyperfine interaction strength
in the principal basis is $C_{\parallel,\text{nn}} = 199$MHz and $C_{\perp,\text{nn}}=123$MHz
while in the NV-basis this corresponds to $C_\parallel(\theta_\text{nn})=129MHz$, 
$C_\perp(\theta_\text{nn})=155MHz$, $C_\text{R}(\theta_\text{nn})=-35MHz$ and 
$C_\Delta(\theta_\text{nn})=25MHz$ ('nn' stands for 'nearest neighbor').
Numerical \emph{ab-initio} calculations found two different classes of third-neighbor positions 
showing a strong hyperfine coupling~\cite{gali-08prb155206}: planar (out of plane) third neighbors 
(see Figure~\ref{fig:diamond}) with coupling constants of $C_{\parallel,\text{3rd}} = 19$ (18) MHz
and $C_{\perp,\text{3rd}}=14$ (13)MHz. 
In ensemble measurements~\cite{felton-09prb075203}, hyperfine ESR lines associated w. third 
neighbors have been identified showing interaction strengths of $C_\parallel^\text{3rd}=18.5$
and $C_\perp^\text{3rd}=13.26$ which is  right in between the theoretically 
predicted values. We use these latter values  as the best estimate of third-neighbor interaction 
strength.

% Level structure
In comparison to the bare NV center, the level structure of the Hamiltonian~\eqref{eq:HNVC} shows
a much larger splitting of the $m_S=\pm 1$ levels due to the much stronger parallel hyperfine interaction
for both carbon positions we considered.
There are two avoided crossings, one strain-avoided at $B_\text{str, nn}=C_\parallel / 2\gamma_e \approx 2.6$mT 
($B_\text{str, 3rd}=0.28$mT for third neighbors) and the other (mainly) exchange-avoided at 
$B_\text{x}=D/\gamma_e \approx 103$mT. 

\begin{figure}
  \hspace{-.45\textwidth} a) \\
   \includegraphics[width=.43\textwidth]{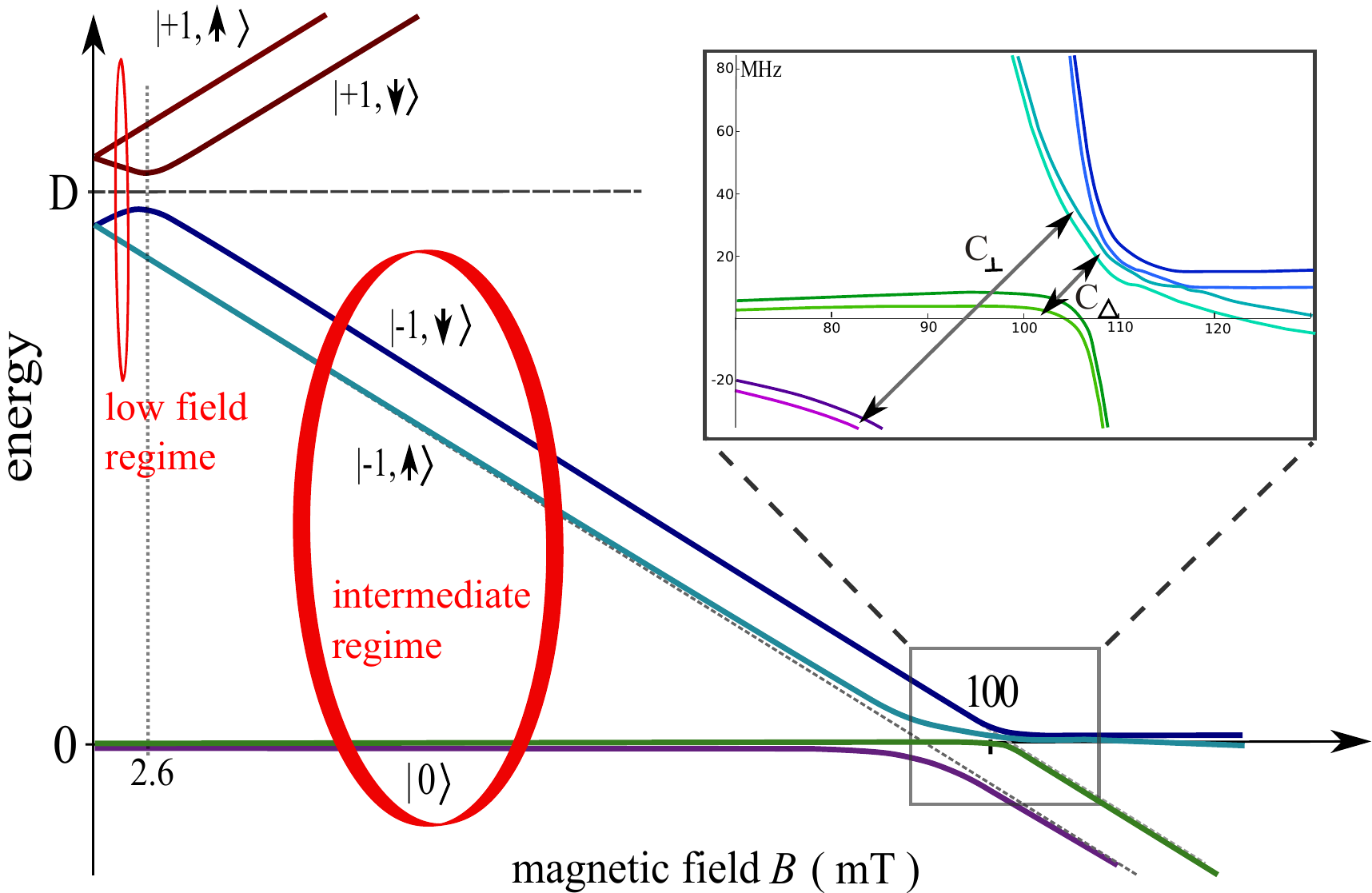} \\
  \hspace{-.45\textwidth} b) \\
  \includegraphics[width=.44\textwidth]{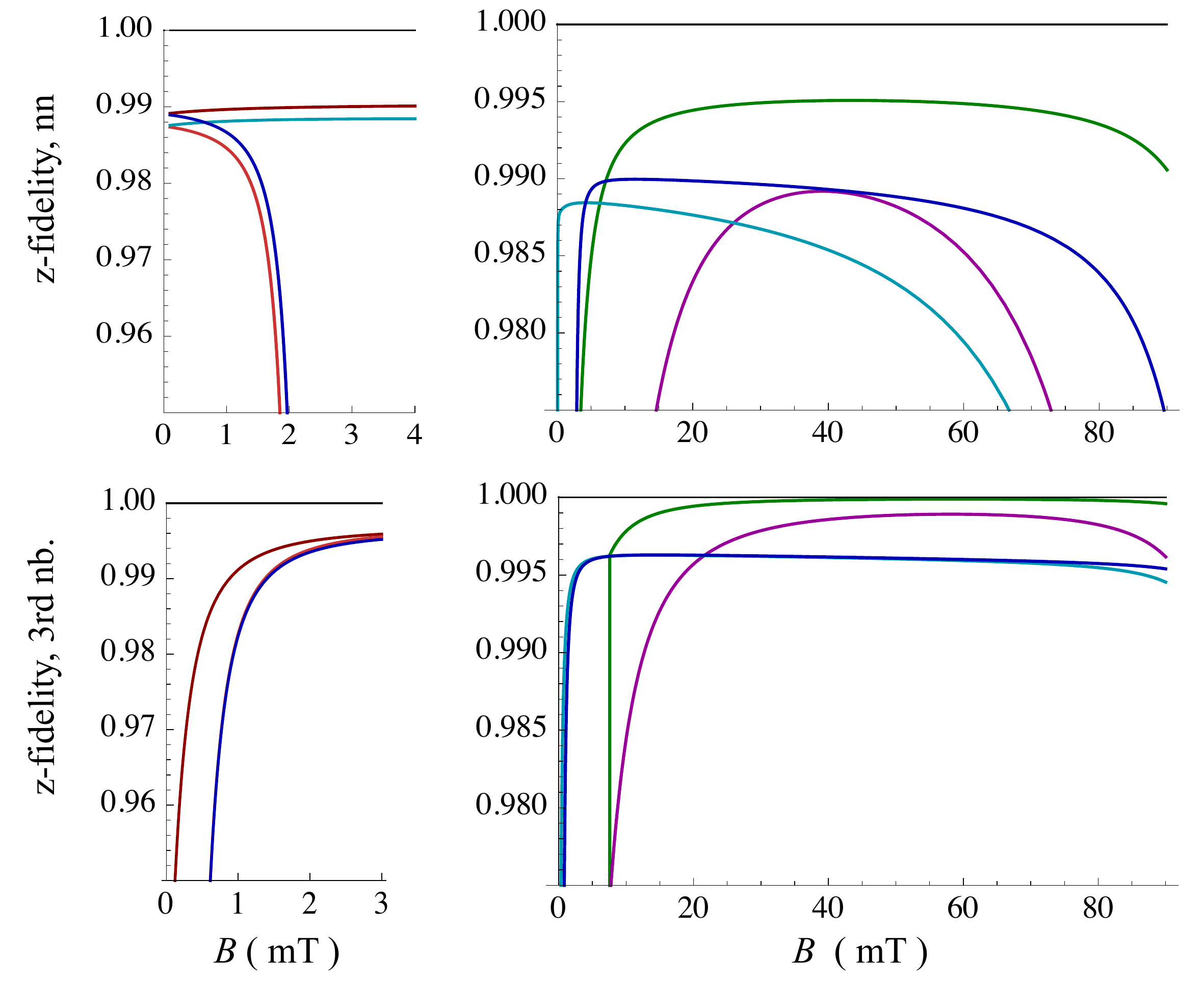}
   \caption{\textbf{ Levels and z-fidelity dependence on  an axial magnetic field.}
   		a.) Energy levels as function of magnetic field strength for nearest neighbor $^{13}$C.  
		 The insets zoom in on the avoided crossings at $B=2.6$mT and 103mT respectively.
		 For this plot, an unrealistically high strain was assumed in order to clearly show the former. 
		 b.) Fidelity of eigenstates with $S_z$,$I_z$-basis ('z-fidelity') for low- and intermediate 
		       magnetic field (left and right column respectively). 
		 Same colors represent the same states in both pictures.
		}
    \label{fig:nvclevels}
\end{figure}

% field strength discussion
For the sake of simplicity in both analysis and application, it makes sense to investigate the model in
magnetic field regimes where the eigenstates have high 'z-fidelity', \ie are close to the $S_z$-$I_z$-basis.
In the NV$^-$, in principle three such regimes exist. The z-fidelity can be achieved for very high magnetic
fields of $B\gg B_\text{x}$, for which the $\ket{m_S=-1}$ levels are lowest in energy. Such large
magnetic fields are however not very desirable from a practical point of view, as they are difficult to keep
stable and the fast Larmor precession of the electronic spin makes accurate timing harder. 
We therefore chose to concentrate on the low field and intermediate field strengths, which are around
$B=1-2$mT and $B=15-50$mT respectively. For nearest neighbor $^{13}$C, this is on either side
of the strain avoided crossing between $\ket{+1}_\text{V}$ and $\ket{-1}_\text{V}$ levels at 
$B=B_\text{str, nn}$ while for third neighbor carbons, both are above $B_\text{str, 3rd}$.

\subsection{Decoherence Model}
% decoherence model
To simulate dissipative time evolution in our system, we solve a time-dependent master equation
with Lindblad operators describing relaxation and dephasing for each subsystem individually 
(the details can be found in Appendix~\ref{app:model})
In order to model the experimentally well established gaussian dephasing of the vacancy spin~\cite{hanson-06prl087601,wang-12prb155204},
we assumed time dependent rates $\gamma_{2,\text{V},a/b}(t) = t / (T_\text{2,\text{V}}^*)^2$ (i.e. same  for 
both dephasing channels a and b).
Since the hyperfine coupling is quite strong for close-by carbons, one should in general use 
Lindblad operators adapted to the eigenbasis of the total system. However, since we are only 
interested in magnetic field regimes where the eigenbasis is very close to the computational 
( $S_z$-$I_z$-)basis, the error due to the simplified decoherence model is inconsequential.
The decoherence times we assumed were $T_{1,\text{V}}=10$ms, $T^*_{2,\text{V}}=100\mu$s, $T_{1,\text{C}}=T_{1,\text{N}}=10$s, 
$T_{2,\text{C}}=T_{2,\text{N}}=10$ms. These are conservative estimates, and each individually has already 
been demonstrated or even surpassed in experiment~\cite{Kennedy-03apl4190, Jelezko-04prl076401,Bala-09nm383}. 

\subsection{Driving}
% driving Hamiltonian, square pulse
We model microwave (MW) and radio-frequency (RF) driving with a Hamiltonian of the form
\begin{equation}
  H_\text{drive} =  u(t) \left(S_x + \frac{\gamma_\text{C}}{\gamma_e} I_{\text{C},x} 
                          + \frac{\gamma_\text{N}}{\gamma_e} I_{\text{N},x}  \right).
    \label{eq:hdrive}
\end{equation}
where the driving field is a sum square pulses $u(t) = \sum_{n=1}^{N_f} \Omega_{0,n} \cos(\nu_n t+\phi_0)$.
The number of frequency components, $N_f$, was in practice either $1$ or $2$ and $\nu_n$ 
usually chosen in resonance with some transition. 
This leaves the $\Omega_{0,n}$ as the main parameter(s) to be optimized. However, we limited 
our search to values which are still in the RWA regime, so that the relative phase $\phi_0$ 
provides control of the driving axis and a direct handle (direct coupling to $y$-direction operators) 
is unnecessary. 

In this work we do not consider pulse shaping (varying $\Omega_0$ and $\phi_0$ continuously in time), 
leaving this as a further optimization to achieve fully fault-tolerant quantum computation in the future.

\section{Low field}
In this section, we present our results for the low magnetic field regime.
As mentioned in the previous section, low magnetic fields offer the advantage of less stringent pulse 
timing requirements. Furthermore, in a scenario where one would like to set up entanglement between
the vacancy and a nuclear spin in the former's $\ket{m_S=\pm 1}$ subspace and then transfer this
bond to a photon via laser excitation of the vacancy, both levels cannot be split by more than the 
laser pulse's line width of $\approx 100$MHz for a short 10ns pulse. Therefore, the magnetic field 
strength values we settled for are a trade-off between the z-fidelity of the eigenstates on one side 
and limiting level separation on the other.  
They are $1.1$mT for nearest- and $2$mT for third neighbor carbon.

\begin{figure}
\hspace{-.4\textwidth} a) \\
\vspace{.3mm}
\includegraphics[width=.475\textwidth]{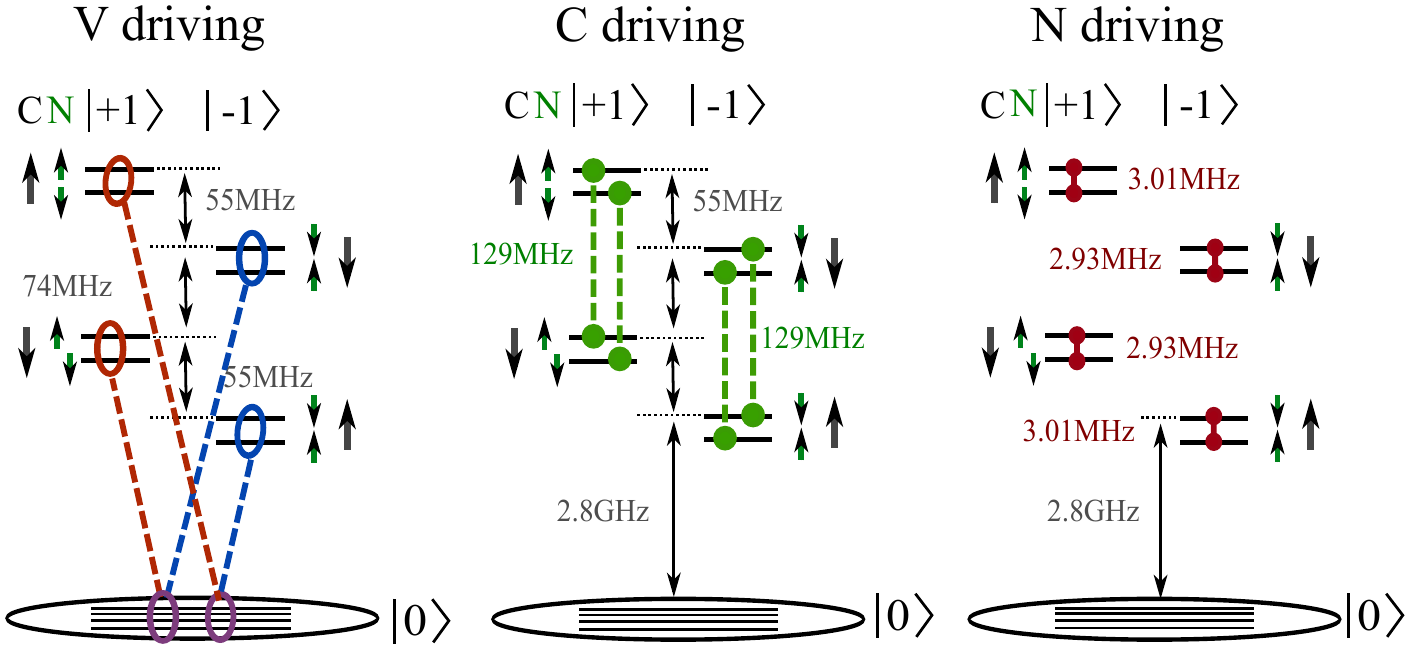} \\
\vspace{.5mm}
\hspace{-.25\textwidth} b) \\
\includegraphics[width=.175\textwidth]{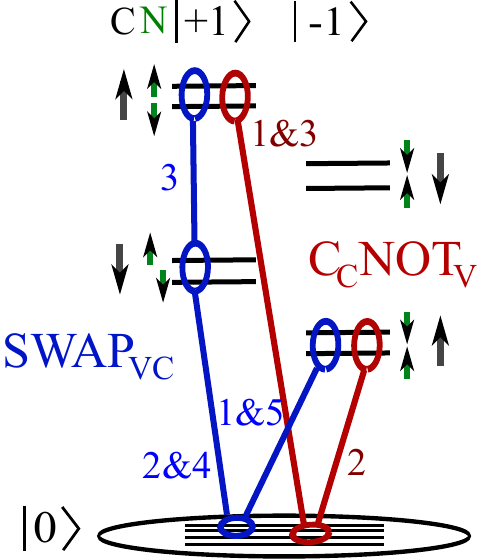}
  \caption{ \textbf{Transitions driven to obtain low $B$ gates.} Level schematics and transition frequency values are for
  	         nearest neighbor carbon at $B=1.1$mT.
  	         (a) Primitive transitions. Controlling the vacancy spin independent
	               of the state of the carbon $^{13}$C requires dual frequency pulses.
		(b) Two-qubit gates in the subspace $\{\ket{+1}_\text{V}, \text{-1}_V \}$
		      require concatenated pulses.  
		}
  \label{fig:lowBgates}
\end{figure}

\begin{figure*}
 \hspace{.125\textwidth} \textbf{nearest neighbor} \hspace{.27\textwidth} \textbf{third neighbor} \\
  \hspace{-.30\textwidth}a) \hspace{.425\textwidth} b) \\
  \parbox{.1\textwidth}{ \vspace{-.2 \textheight} {\large V} \\ $\ket{0}\rightarrow\ket{+1}$} \includegraphics[width=.4075\textwidth]{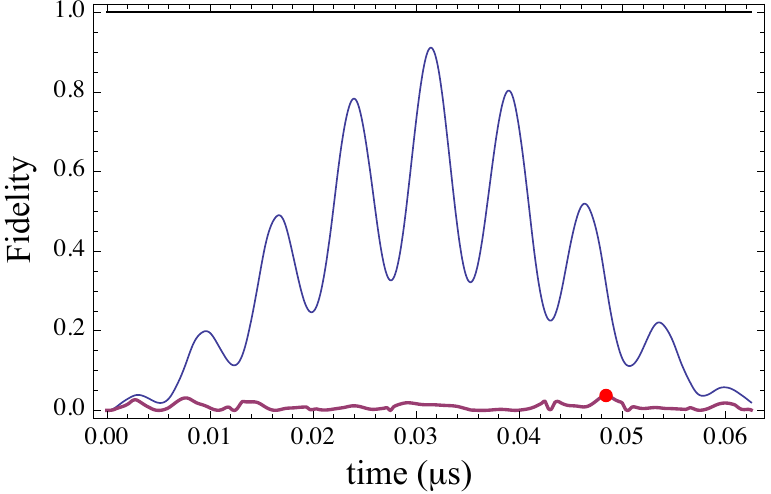}\hspace{.05\textwidth} \includegraphics[width=.385\textwidth]{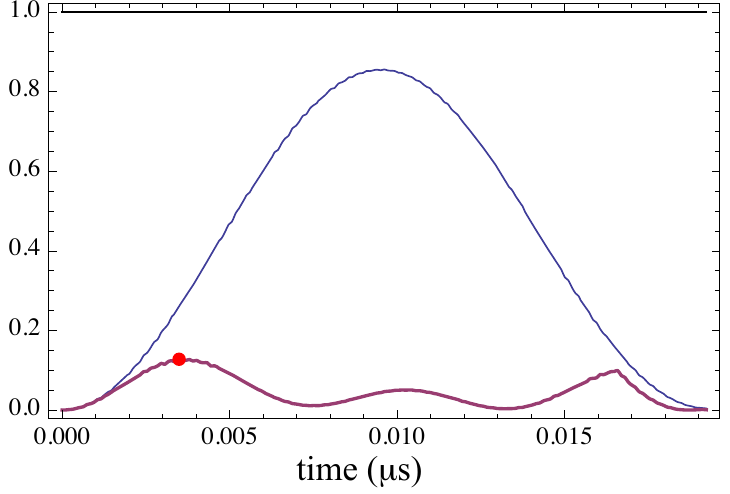} \\

  \vspace{.01\textheight}
      \hspace{-.30\textwidth}c) \hspace{.425\textwidth} d) \\
   \parbox{.1\textwidth}{ \vspace{-.2 \textheight} {\large V} \\ $\ket{0}\rightarrow\ket{-1}$}  \includegraphics[width=.4075\textwidth]{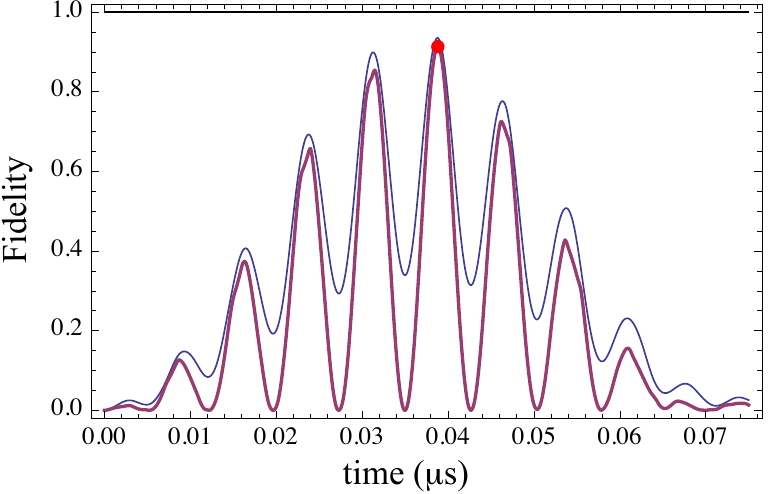}\hspace{.05\textwidth} \includegraphics[width=.385\textwidth]{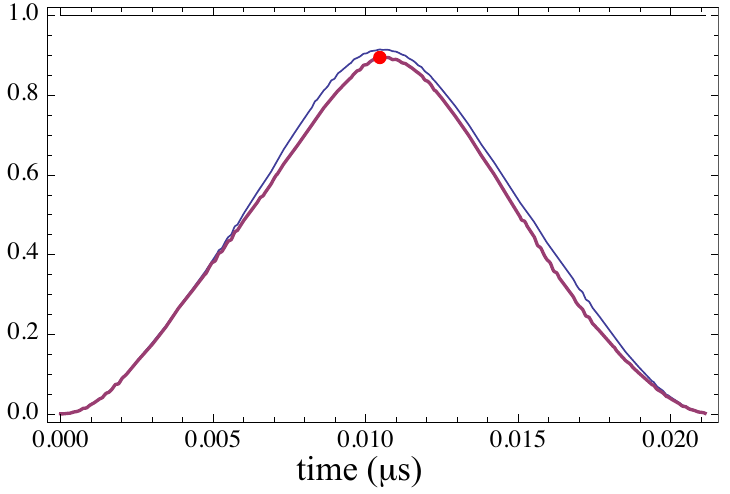} \\
    \vspace{.01\textheight}
      \hspace{-.30\textwidth}e) \hspace{.425\textwidth} f) \\
   \parbox{.1\textwidth}{ \vspace{-.2 \textheight} {\large C} \\ $\ket{\up}\rightarrow\ket{\dn}$}  \includegraphics[width=.4075\textwidth]{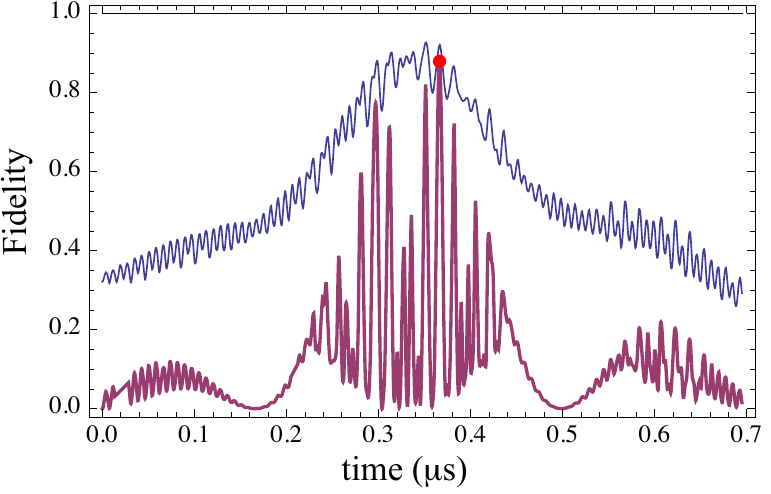}\hspace{.05\textwidth} \includegraphics[width=.385\textwidth]{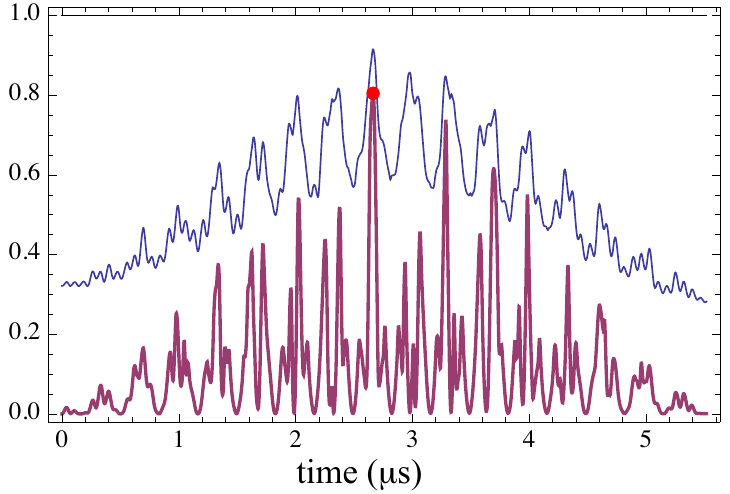} \\
  \caption{  Low-$B$ $\pi$-pulse gate fidelities for the carbon in nearest neighbor (left column) and 
                    third neighbor positions (right column).
                    The first column refers to the $\ket{0}_\text{V}\rightarrow\ket{+1}_\text{V}$ transition
                    the second to  $\ket{0}_\text{V}\rightarrow\ket{-1}_\text{V}$ and the last to
                     $\ket{\up}_\text{C}\rightarrow\ket{\dn}_\text{C}$.
                    The purple trace is the gate fidelity, with the average state fidelity shown in blue
                    as a reference. For the $m_S=+1$ states, individual state pulses are out of phase,
                    resulting in very low gate fidelity, while average state fidelities are similar to
                    the ones for $m_S=-1$. 
  		}
  \label{fig:vdrive}
\end{figure*}

% short version 
\begin{table*}
  \begin{tabular}{lc|c|c|c|c|c|c|}
      transition                                    &                          other     &    \multicolumn{3}{c|}{nearest neighbor}                                           &  \multicolumn{3}{c|}{third neighbor}  \\                  
                                                          &                                         &  $\Omega_0^\text{opt}$(MHz) &  fidelity (\%) & time $T_\pi$ (ns)   &  $\Omega_0^\text{opt}$ (MHz) &  fidelity (\%)  & time $T_\pi$ (ns)  \\
      \hline
      $\ket{0}_\text{V}\rightarrow\ket{+1}_\text{V}$ & $\ket{\up,\up}_\text{CN}$ &  36 ( 45 )   &  98.5 ( 99.0 )   &  19.1 ( 16.2 )     &     32 ( 43 )      &  98.3 ( 99.0 )       &  21.1 ( 17.2 )     \\
     $(\,\ket{-1}_\text{V})$                 &  $\ket{\up, x+}_\text{CN}$                          &   66 (  "  )     &  97.1 ( 97.6 )   &  15.7 ( 15.9 )     &     48  (  "  )      &  99.0 ( 99.1 )       &  14.7 ( 17.4 )     \\
			                               &              $\ket{x+, \up}_\text{CN}$              &   50 (  "  )     &  97.2 ( 98.0 )   &  16.0 ( 16.1 )     &     75 (  77  )    &  90.0 ( 91.2 )       & 8.9 ( 9.9 )        \\ 
			                               &              $\ket{x+, x+}_\text{CN}$                &  24 ( 22.5 ) &  93.2 ( 94.2 )   &  31.0 ( 31.1 )     &     74 (  76  )    &  89.3 ( 90.1 )       &  9.0 ( 9.9 )        \\ 
	&  $(\frac{1}{2}\sigma_0)_\text{C}\otimes(\frac{1}{2}\sigma_0)_\text{N}$ &   " (  "  )     &  95.3 ( 95.7 )   &  31.8 ( 31.5 )     &   " (  "  )            &  94.1 ( 94.4 )       &  9.3 ( 10.5 )  \\
					            \cline{2-8}
					            &       gate                                              &  22.5(19.0)	  &  4 (91.3)\%  & 48.4 (38.8)                   & 70 ( 70 )	    &  13.4\% (89.2\%)  &  3.9 (10.4)  \\
					            &       avg                                               &  		            &  90.6 (94.3)\%    &  32 (38.7)             &  "    (  " )	    &   87.1\% (91.3\%) &  9.8  ( " )  \\
        \hline
 $\ket{\up}_\text{C} \rightarrow \ket{\dn}_\text{C}$ & $\ket{+1, \up}_\text{VN}$ &  70    &   99.2   &  322         &     100    &   99.3   &  1654  \\
 					 				    & $\ket{+1, x+}_\text{VN}$ &   61    &   99.1   &  335         &       35     &   98.5   &  4649  \\	
		      & $\frac{1}{\sqrt{2}}(\ket{+1}+\ket{-1})_\text{V}\ket{\up}_\text{N}$ &    "    &   97.4   &  312        &       99     &   95.2   &  1724  \\
					             & $ " \quad (\frac{1}{2}\sigma_0)_\text{N}$   &    "    &   98.2   &  328         &        "       &   92.2   &  1667  \\
					            \cline{2-8}
					            &      gate                                                     &  51   & 87.9\%  & 367                 &  61          & 80.5\%  & 2665 \\
				                     &      avg                                                       &         & 91.8\%   & "                      &                 & 90.8\%  &   "   \\
     \hline
      $\ket{\up}_\text{N} \rightarrow \ket{\dn}_\text{N}$ & $\ket{+1, \up}_\text{VC}$ &  100    &   97.9   & 16900   &  135    &   94.1   & 6747  \\
 					 				    & $\ket{+1, x+}_\text{VC}$       &   119   &   63.6   &  7025     &   122    &   89.6  &  7043  \\
			& $ \frac{1}{\sqrt{2}}(\ket{+1}+\ket{-1})_\text{V}\ket{\up}_\text{C}$ &    "    &   74.2   & 11600      &  111   &   91.4   &   9134  \\		
     \hline	      	                               
  \end{tabular}
  \caption{Maximum fidelities for square pulse driving of various specific starting states as well as gate fidelities 
  		for both carbon positions. 
		Gate fidelities refer to the quantity
		 $\underset{t}{\text{max}}\, \underset{j}{\text{min}}\, \text{Fid}(\mathcal{E}_t[\rho_{0,j}] , \rho_{\text{T},j})$
		 where $\mathcal{E}_t$ is the time evolution operator and $j$ runs from 1 to the number of initial-target state 
		 pairs (25, 16 and 8 for vacancy, carbon and nitrogen).
  		}
  \label{tab:lowB}
\end{table*}

\subsection{Single qubit gates}
The pulses and pulse sequences needed for single-qubit control are illustrated in
Figure~\ref{fig:lowBgates}. In the following, unless otherwise stated, fidelities and times 
given apply to a single $\pi$-pulse. We also want to distinguish between state-driving 
fidelity and gate fidelity: the former refers to the fidelity 
$F_{\rho_0} \!= \! F\left(\mathscr{E}_T[\rho_0], \rho_\text{target} \right)$ between 
the time evolution of one particular starting state and its intended target state,
where $\mathscr{E}_T[\rho]$ is the superoperator describing the time evolution of a
density matrix $\rho$ until time $T$ and $F(.,.)$ is the fidelity measure as described
in Appendix~\ref{app:fidmeas}.
Here we usually have $\rho_\text{target} = U_\text{id} \rho_0 U_\text{id}^\dagger$ with $U_\text{id}$ some
desired unitary operation.
Gate fidelity is then the minimum of $F_{\rho_0}$ over the entire Hilbert space of our system: 
$F(\mathscr{E},U_\text{id}) = \underset{\ket{\psi}\in\mathcal{H}}{\text{min}}\, F_{\ket{\psi}\bra{\psi}}$.
This is hard to compute exactly even for our modest Hilbert-space dimension of
 $\text{dim}\,\mathcal{H}_\text{NV+C}=12$.
Therefore we settled for an approximation by sampling the Hilbert space at representative points. 
For a detailed description of how we measure fidelity in our numerical implementation we refer 
to Appendices~\ref{app:fidmeas} and~\ref{app:model}.

\emph{Driving V. ---}
The transition frequency between $\ket{+1}_V$ and $\ket{-1}_V$ is strongly dependent on the
state of the carbon nuclear spin for both nearest and 3rd nearest neighbor $^{13}$C, which
clearly poses a problem for single-qubit operations. 
We had to solve this in two different ways for the two carbon positions: in the case of nearest
neighbor using dual frequency driving ($N_f=2$ in~\eqref{eq:hdrive}) works well, while it does
not give good results for third neighbors. 
We attribute this to the much stronger parallel hyperfine interaction in the former case, resulting
in a splitting of  $\approx C_\parallel^\text{nn} =129$MHz between carbon $\ket{\up}$ and $\ket{\dn}$.
This is resolvable within the $\pi$-pulse times giving the best fidelities, which are on the order of 
O(10ns). 
In contrast, the splitting is only $\approx 13.5$MHz for third nearest neighbor and therefore not big enough
to allow resolution of the two-component pulse within a time of about 10ns. This would rather require 
one order of magnitude longer pulses \ie weaker driving power. Unfortunately we found that for
such slow pulses maximum fidelity invariably suffers. The best solution in this case is then
to apply a fast pulse tuned to the average transition frequency. In principle it holds: the faster
the better, but for very short pulse times, timing error will start to seriously reduce the fidelity.

When starting in a polarized state, we find that state fidelities can reach up to 98.5\% 
for nearest neighbor ($\ket{\psi_0} = \ket{-1,\up,\up}_\text{vcn}$, $\Omega_0=45$MHz, $\pi$-time 
of $T=16.2 $ns) and 98.3\% for third neighbor  carbon 
($\ket{\psi_0} = \ket{-1,\up,\up}_\text{vcn}$, $\Omega_0=30$MHz, $\pi$-time of $T=25.0 $ns, see 
Table~\ref{tab:lowB}).
Gate fidelities are significantly lower. In fact, transitions from $m_S=0$ to $m_S=+1$ show an
intriguing disconnect between average and gate fidelity: average fidelity reaches about 90\%,
similar to $m_S=-1$ transitions. Gate fidelity, as defined above, is however only around 10\% 
or less, showing that the $\pi$-pulse times for the individual starting states must be very different 
('out of phase'). This is not the case for the transitions between $m_S=0$ and $m_S=-1$, where 
the gate fidelity reaches within 2-3\% of the average state driving fidelity.

These fidelities are all for single $\pi$-pulses. Single qubit gates in the physical basis 
$\{\ket{+1}_\text{V},\ket{-1}_\text{V} \}$ require three consecutive pulses and will thus have 
lower fidelity still.
In general, fidelities depend on driving power $\Omega_0$, but for V  this is not as pronounced
as for the two nuclear spins.

\emph{Driving C. ---}
The level spitting for the carbon is independent of either the state of V or N, thus manipulate the
carbon spin state independently. 
This probably explains why it shows the highest state fidelities of the three subsystems, 
reaching 99.2\% (99.3\%) for nearest (third) neighbors (cp. Table~\ref{tab:lowB} ) if
vacancy and nitrogen spins are polarized. Gate fidelity is much lower however, with
88\% and 80.5\% for nearest and 3rd nearest neighbors respectively.

\emph{Driving N. ---}
Similar to the vacancy spin, transition frequencies for the nitrogen nuclear spin depend on the 
state of the carbon, with a difference between level splittings of $\omega_{N,\up\dn}\approx 800$kHz
between $\ket{\up}_C$ and $\ket{\dn}_C$.
This means that while the vacancy spin can in principle be in an arbitrary state, C must be polarized 
to either $\up$ or $\dn$. This is in itself somewhat remarkable, since in our model we have no 
direct coupling between the nuclear spins.
The maximum fidelity is 97.9\% for the starting state $\ket{+1,\up}_{VC}$ while gate fidelity is much lower,
mostly due to the energy-splitting difference mentioned as well as drift of the carbon spin phase. 

A summary of the results for nearest neighbor and third nearest neighbor carbon is given in 
Table~\ref{tab:lowB}.
 
\subsection{Multi-qubit gates and entanglement}
\emph{Driven gates. ---}
As we mentioned before, the hyperfine interaction causes transitions for the vacancy and nitrogen to be 
dependent on the state of the other qubits. While this is a problem when implementing single-qubit gates,
it can be used to implement two-qubit gates via driving.
Using a qubit basis consisting of $\ket{m_S=0}$ and either of $\ket{m_S=\pm 1}$ such
gates can be implemented with a single pulse. 
For the basis $\{\ket{+1}_\text{V}, \ket{-1}_\text{V}\}$ there is the difficulty that direct transition between 
these levels are not dipole-allowed and therefore exceedingly slow when driven directly.
Thus, between the $\ket{m_S=\pm 1}$ states, all two-qubit gates involving V must be realized via
sequences of at least three entangling pulses plus single-qubit rotations to tidy up factors of $i$.
Figure~\ref{fig:lowBgates}b shows two examples for such gates. 

For example a CNOT$_\text{C,V}$ (logical $\ket{0}_\text{C}$ corresponds to physical $\ket{\dn}_\text{C}$) 
 would consist of the sequence
$\pi(\ket{+1,\up}_\text{VC}\leftrightarrow \ket{0,\up} ), \pi(\ket{0,\up}\leftrightarrow \ket{-1,\up} ), 
  \pi(\ket{0,\up}\leftrightarrow \ket{+1,\up} )$. To be independent of the nitrogen, the pulse times 
  must be fast compared to the nitrogen hyperfine level splitting of $3$MHz (=330ns), but slow 
  enough to minimize off-resonant driving of the wrong transition (to $\ket{m_S=-1}$). 
  For CNOT$_\text{C,V}$ the two transitions are separated by about $180$MHz at $B=1$mT
  corresponding to roughly 6 ns.  Thus, both criteria can only be satisfied to limited
  degree, with the ideal pulse length being about 45ns per pulse or 135ns in total. 
  A SWAP gate  between vacancy and carbon state requires 5 $\pi$-pulses (see Figure~\ref{fig:lowBgates}b) 
  and has thus a lower fidelity still.
 
 \emph{Entangled states. ---}
 As we have seen, implementing multi-qubit gates with high fidelities is difficult in the low 
 magnetic field regime.  However, if we aim for something less ambitious, such as preparing some 
 useful state from a known starting state, high fidelities are achievable even when including the 
 nitrogen.
 As examples, let us look at two entangles states $\left(\ket{+1,\dn}+\ket{-1,\up}\right)_\text{VC}/\sqrt{2}$ and
  $\left(\ket{+1,\dn}+\ket{-1,\up}\right)_\text{VN}/\sqrt{2}$. 
 The standard way to reach the former is the three-pulse sequence 
 $\pi/2(\ket{0,\dn}_\text{VC}\leftrightarrow \ket{-1,\dn} ), \pi(\ket{-1,\dn}\leftrightarrow \ket{-1,\up} ), \pi(\ket{0,\dn}\leftrightarrow \ket{+1,\dn} )$
 involving only 'allowed' (=single flip) transitions. Similalry, for the latter state we would have
 $\pi/2(\ket{0,\dn}_\text{VC}\!\leftrightarrow\! \ket{-1,\dn} ), \pi(\ket{-1,\dn}\!\leftrightarrow\! \ket{-1,\up} ), \pi(\ket{0,\dn}\leftrightarrow \ket{+1,\dn} )$,
 in both cases assuming a starting state $\ket{0,\dn,\dn}$. For these sequences we find maximum 
 fidelities of 97.4\% (97.3\%). 
 However, the presence of the strong hyperfine interaction makes it possible to directly drive 
 ordinarily 'forbidden' transitions involving two simultaneous spin flips. This lets us reach the 
 target states with the two-pulse sequences
 $\text{MW-}\pi(\ket{0,\dn}_\text{VC}\leftrightarrow\ket{-1,\dn}), \text{RF-}\pi/2(\ket{-1,\dn}_\text{VC}\leftrightarrow\ket{+1,\up})$
 and
 $\text{MW-}\pi(\ket{0,\dn}_\text{VN}\leftrightarrow\ket{+1,\dn}), \text{RF-}\pi/2(\ket{+1,\dn}_\text{VN}\leftrightarrow\ket{-1,\up})$.
 For these fidelities are 98.5\% and 98.9\% at optimum gate times of 286ns (15$\mu$s), significantly 
 better than for the ordinary three pulses.
 This two pulse scheme works well only for setting up odd-parity Bell states, because the 
 two-spin flip processes are mainly caused by the hyperfine exchange term. Even-parity Bell 
 states would need a counter-rotating term, which is only present for the carbon nuclear spin 
 and there too it is much weaker than the exchange term.

\section{Intermediate field}
 For magnetic field strengths between $B\approx 15$mT and $50$mT the eigenstates are much
 closer to the  $S_z$-$I_z$-basis than for low $B$ (see Section II, Fig.~\ref{fig:nvclevels} b)). 
 If we choose the $\ket{m_S= 0}$ and $\ket{-1}$ levels as our vacancy-qubit basis, we see that 
 while the z-fidelity of some states reaches a maximum only much later, there average peaks in the 
 region around 25mT and this is therefore the value we choose. It has the added benefit of large 
 detuning with and thus low leakage into the $\ket{m_S=+1}$ subspace, which has to be avoided
 as it would constitute a qubit loss error.

\subsection{Single qubit gates}
\begin{figure}
  \includegraphics[width=.31\textwidth]{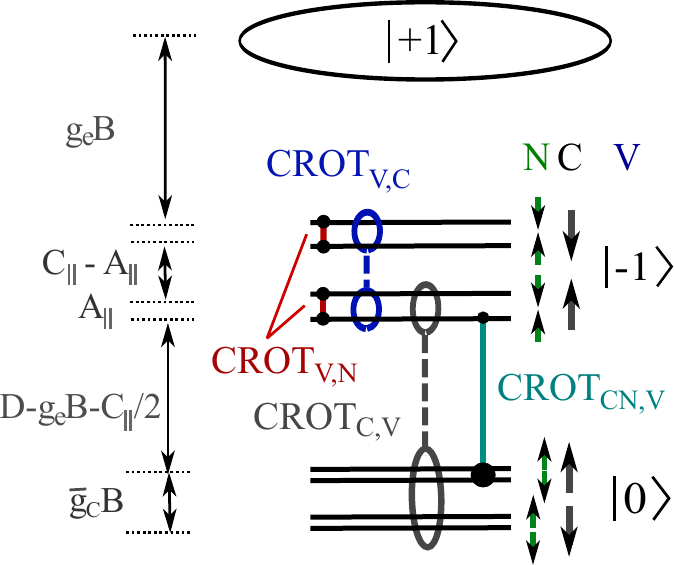}
  \caption{Energy level schematic with transitions yielding multi-qubit gates. 
		CROT$_(c[c'],t)$ stands for controlled rotation of qubit t by qubit(s) c (and c'). 
		They are equivalent to a CNOT for one and a TOFFOLI gate for two control qubits .
		} 
   \label{fig:medBgates}
\end{figure}

\begin{figure*}
 \hspace{.125\textwidth} \textbf{nearest neighbor} \hspace{.27\textwidth} \textbf{third neighbor} \\
  \hspace{-.30\textwidth}a) \hspace{.425\textwidth} b) \\
  \parbox{.1\textwidth}{ \vspace{-.2 \textheight} {\large V} \\ $\ket{0}\rightarrow\ket{-1}$} \includegraphics[width=.4075\textwidth]{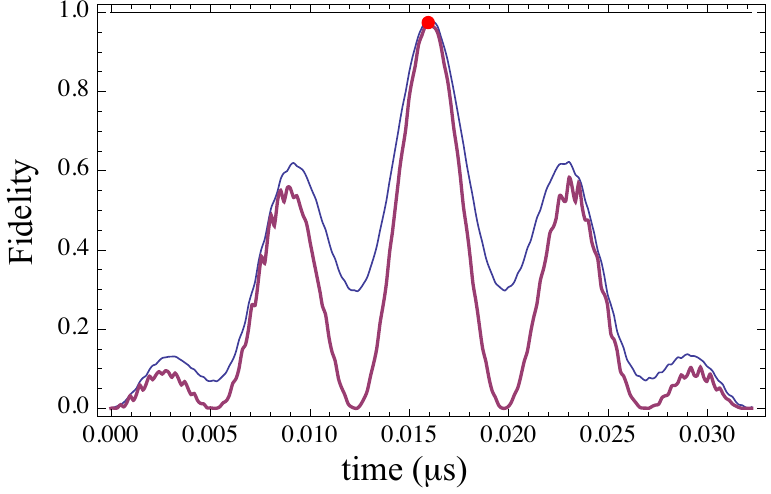}\hspace{.05\textwidth} \includegraphics[width=.384\textwidth]{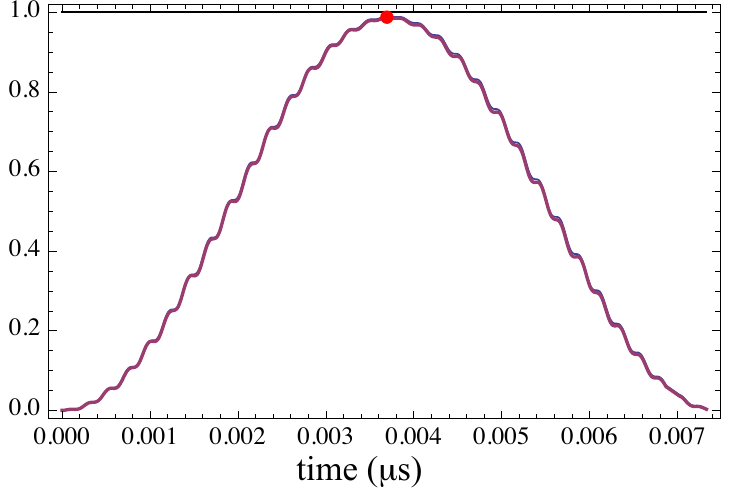} \\
  \vspace{.01\textheight}
      \hspace{-.30\textwidth}c) \hspace{.425\textwidth} d) \\
   \parbox{.1\textwidth}{ \vspace{-.2 \textheight} {\large C} \\ $\ket{\dn}\rightarrow\ket{\up}$}  \includegraphics[width=.4075\textwidth]{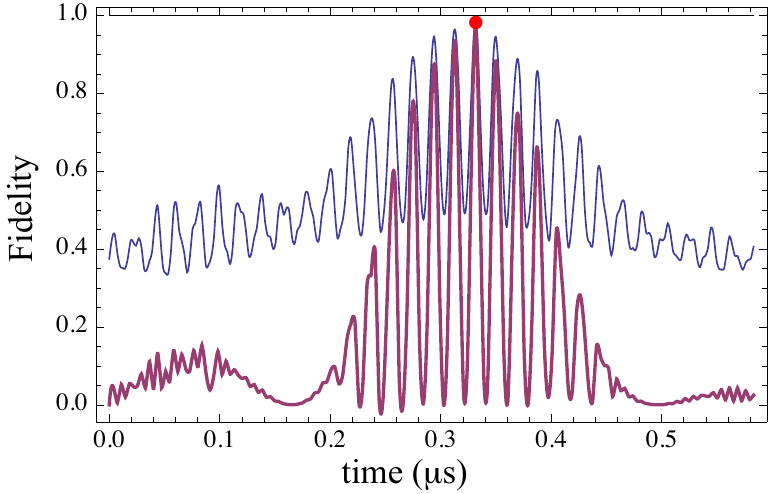}\hspace{.05\textwidth} \includegraphics[width=.384\textwidth]{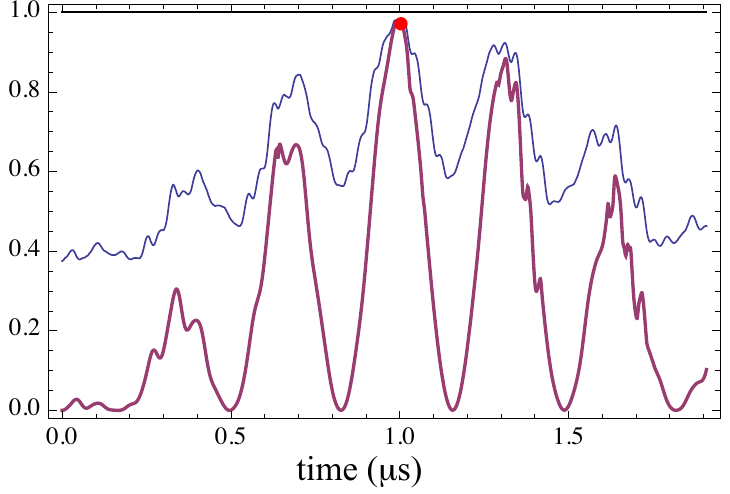} \\
    \vspace{.01\textheight}
      \hspace{-.30\textwidth}e) \hspace{.425\textwidth} f) \\
   \parbox{.1\textwidth}{ \vspace{-.2 \textheight} {\large N} \\ $\ket{\dn}\rightarrow\ket{\up}$}  \includegraphics[width=.4075\textwidth]{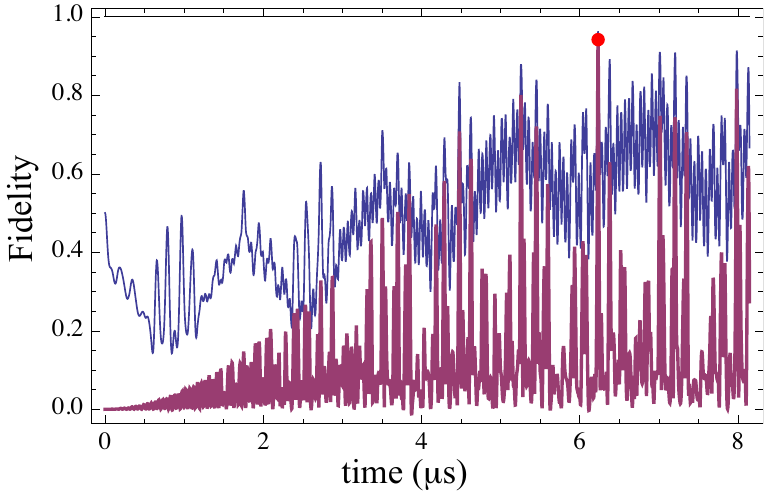}\hspace{.05\textwidth} \includegraphics[width=.384\textwidth]{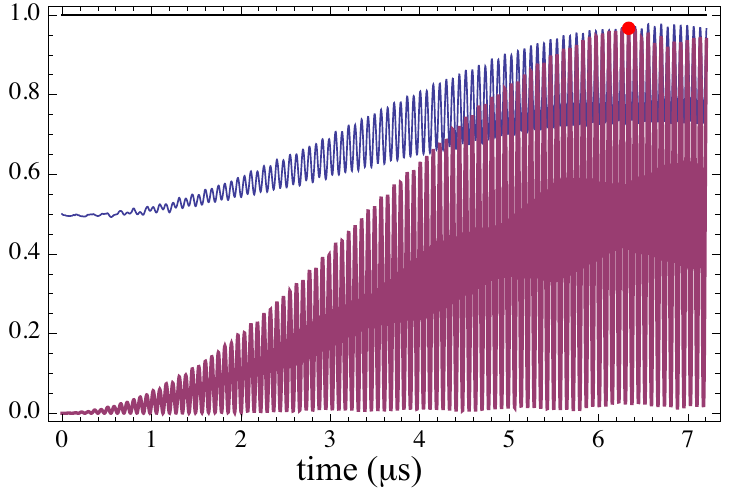} \\
  \caption{ Gate fidelities of vacancy electron $\pi$-pulse driving at low $B$-field for nearest neighbor (a,c,e)
  		and third nearest neighbor carbon (b,d,f). The horizontal axis shows time in $\mu$s and
		the vertical axis fidelity. Thick, purple traces show the gate, and the thin blue traces average
		state-driving fidelity. 
		The insets show a zoom-in on each maximum.
  		}
  \label{fig:medBgatefids}
 \end{figure*}

\emph{Driving V. ---}
 From the energy level structure at intermediate B (Figure~\ref{fig:medBgates}) one sees, that like
 in the low-B regime, controlling the vacancy independent of the carbon spin state is again not 
 straightforward. 
 As before, our solution was the dual-frequency driving technique in case of nearest neighbor 
 carbon and driving the average transition frequency in case of third-neighbor carbon.
 With this, we were able to achieve maximum fidelities of 96.1\% and 97.7\% respectively.
 Plots of the gate fidelity for a $\pi$-pulse are shown in Figure~\ref{fig:medBgatefids} for
 both carbon positions.
 Naturally, state fidelities are higher, up to 99.3\% (99.7\%) when the nuclear spins are polarized
 (see Tables~\ref{tab:medB}).

\emph{Driving C. ---}
 Unlike at low magnetic field, fidelities of the carbon nuclear spin nearly match those
 for the vacancy. 
 Our choice of computational basis means however, we can only effect a $\pi$-rotation, if the 
 vacancy spin is polarized into logical $\ket{1}_\text{V}$ (the $\ket{m_S=-1}$ state). In our 
 numerical gate fidelity computations we nonetheless included starting states with $\ket{0}_\text{V}$, 
 in which case we checked how well the pulse preserves this state, \ie we set target state 
 equal starting state for the gate fidelity estimation.
 For nearest neighbor carbon, starting states with the vacancy spin polarized show fidelities 
 up to 99.6\% while falling of somewhat if the vacancy starts in the state
 $\ket{x+} = (\ket{0}_\text{V} + \ket{-1}_\text{V})/\sqrt{2}$. 
 There is no significant difference between these starting states in case of third neighbor
 carbon.
 
\emph{Driving N. ---}
 Compared to the low magnetic field regime, nitrogen transition frequencies depend far less 
 on the state of the carbon, which allows relatively good gate fidelities of 96.6\% for
 third nearest neighbor and 94.1\% for nearest neighbor $^{13}$C.
 With gate times on the order of 6$\mu$s non-polarized states of the vacancy spin would have 
 dephased strongly due to the low assumed electron $T_2$ time of 100$\mu$s 
 (still relatively long for a solid state qubit), which is why excluded them from our
 gate fidelity computation.
 Since relaxation is much slower polarized vacancy spin states do not suffer appreciably 
 during the gate time and thus state fidelity for such starting states is as high as
 for the other subsystems 99.2\% and 99.4\% respectively for the two different
 carbon positions.
 
 All results are summarized in Table~\ref{tab:medB} for nearest 
 and third neighbor carbons respectively. 
 
 %short version
\begin{table*}
  \begin{tabular}{lc|c|c|c}
     \hspace{-.04\textwidth}  \textbf{A : nearest neighbor} \\
      principal                                   &                          other                              &  $\Omega_0^\text{opt}$ (MHz) & $\pi$-pulse fidelity (\%) & time $T_\pi$ (ns)   \\
      \hline
      $\ket{0}_\text{V}\rightarrow\ket{-1}_\text{V} $ & $\ket{\up,\up}_\text{CN}$     &    31   			&  $99.3\pm .2$       &  23.4       \\
     ($\nu=$2125 MHz)                &              $\ket{\up, x+}_\text{CN}$                        &    44                         &  $98.5\pm .3$       &  16.0        \\
			                               &              $\ket{x+, \up}_\text{CN}$                    &     "                           &  $98.2\pm 1.3$         &   16.4   \\
					             &              $\ket{x+, x+}_\text{CN}$                     &     "                              & $98.1\pm 1.2$          &  16.0     \\		            
            &  $(\frac{1}{2}\sigma_0)_\text{C})\otimes(\frac{1}{2}\sigma_0)_\text{N}$ &   "  			&  $98.7\pm .3$  	& 16.4  \\
                                                          \cline{2-5}
					            &       gate:                                                                &  44          	         &  $96.1\pm 1.3$  	& 16.0  \\
 	\hline
	CROT$_\text{C,V}$       & $\nu=2123.6$ MHz & 43       &  $96.8\pm 0.3$	 & 15.8 \\	 
	CROT$_\text{CN,V}$    &  $\nu=2122$ MHz   & 0.8       &  $95.2\pm 0.01$    &  932 \\	    
        \hline
 $\ket{\up}_\text{C} \rightarrow \ket{\dn}_\text{C}$ & $\ket{-1, \up}_\text{VN}$       &  31    			&  $99.6\pm .1$   	&  486.4  \\
 ($\nu=126.5$MHz)		   & $\ket{-1, x+}_\text{VN}$  				  	 &   "    			&  $99.6\pm .1$   	&  486.4  \\
				
		          & $\frac{1}{\sqrt{2}}(\ket{0}+\ket{-1})_\text{V}\ket{\up}_\text{N}$ &    32.5    			&  $98.3\pm .4$   &  453.1  \\
					                                      & $ " \quad \ket{x+}_\text{N}$    &    52.5    		&  $98.3\pm .4$   &  331.5  \\
					                 & $ " \quad (\frac{1}{2}\sigma_0)_\text{N}$ &   "   			&  $98.0\pm .4$   &  331.4      \\
					            \cline{2-5}
				                     &  gate:                                                       		&  52.7   			& $98.4\pm .2$  & 332  \\
     \hline
         	CROT$_\text{V,C}$              & $\nu=126.5$ MHz& 52.7  &  $97.9\pm 0.1$ & 325 \\	 
      \hline
      $\ket{\up}_\text{N} \rightarrow \ket{\dn}_\text{N}$ & $\ket{-1, \up}_\text{VC}$ & 110    			&  $99.2\pm .04$      & 6719  \\
  ($\nu=3.13$MHz)		 				    & $\ket{-1, x+}_\text{VC}$       &   "				&  $96.2\pm 1.0$    	& 7146  \\
				             & $ \ket{-1}_\text{V}(\frac{1}{2}\sigma_0)_\text{C}$ &  121   			&  $99.0\pm .01$   	& 6143   \\
                                                      \cline{2-5}	 
           					& gate:  								&     109.5	                  &  $94.1\pm .3$ & 6232 \\
        \hline
           CROT$_\text{V,N}$      & $\nu=3.13$ MHz&  109     &  $91.0\pm .01$	 & 6232 \\	
        \hline
          \\
       \hspace{-.04\textwidth}  \textbf{B : third neighbor} \\
      principal                                   &                          other                             &  $\Omega_0^\text{opt}$(MHz) & $\pi$-pulse fidelity (\%) & time $T_\pi$ (ns)   \\
      \hline
      $\ket{0}_\text{V}\rightarrow\ket{+1}_\text{V} (\,\ket{-1}_\text{V})$ & $\ket{\up,\up}_\text{CN}$ &  72   &  $99.7\pm .2$      &  9.6       \\
      ($\nu=2181$)&   $\ket{\up}_\text{C}\otimes(\frac{1}{2}\sigma_0)_\text{N} $ 	&  90   &  $99.4\pm .2$      &     7.9         \\
					             &              $\ket{x+, x+}_\text{CN}$                     &   129                             & $97.2\pm 1.5$      &     3.0     \\
			&  $\ket{x+}_\text{C}\otimes(\frac{1}{2}\sigma_0)_\text{N}$)       &    230                            & $97.0 \pm 1.6$      &    3.0      \\			            
            &  $(\frac{1}{2}\sigma_0)_\text{C})\otimes(\frac{1}{2}\sigma_0)_\text{N}$ &    "			   &  $97.5\pm 1.6$  	   &   3.0  \\
        						\cline{2-5}
					            &       gate :                         				&  190       	         &  $97.7\pm 0.9$\%  	& 3.7       \\
	\hline
	   CROT$_\text{C,V}$       &$\nu=2173.4$ MHz&  12.5       &  $92.7\pm 0.1$ & 58.0 \\	 
	   CROT$_\text{CN,V}$   &   $\nu=2171.8$ MHz& 1.1  &  $97.4\pm 0.01$ &  634    \\	
        \hline
 $\ket{\up}_\text{C} \rightarrow \ket{\dn}_\text{C}$ & $\ket{-1, \up}_\text{VN}$      &    110    			&  $99.6\pm .02$   	& 1336  \\
 ($\nu=13.45$MHz)		   & $\ket{-1, x+}_\text{VN}$  				  	 &    51    			&  $99.4\pm .01$   	&  2973  \\
		          & $\frac{1}{\sqrt{2}}(\ket{0}+\ket{-1})_\text{V}\ket{\up}_\text{N}$ &     72   			&  $99.3\pm .6$   & 1980  \\
					                                      & $ " \quad \ket{x+}_\text{N}$    &     "    			&  $99.2\pm .11$   & 1981   \\
					                 & $ " \quad (\frac{1}{2}\sigma_0)_\text{N}$ &    72.5   			&  $99.2\pm .10$   & 1980     \\
					            \cline{2-5}
				                     &  gate:                                                       		&  130.5   			&  $96.9 \pm .02$  & 1001 \\
      \hline
         	CROT$_\text{V,C}$              & $\nu=13.5$ MHz& 130  &  $98.2\pm 0.05$ & 1082 \\	
     \hline
      $\ket{\up}_\text{N} \rightarrow \ket{\dn}_\text{N}$ & $\ket{-1, \up}_\text{VC}$ &  83   			&  $99.4\pm .01$    & 8847  \\
  ($\nu=3.08$MHz)		 				    & $\ket{-1, x+}_\text{VC}$       &  101			&  $98.4\pm .05$    	& 7649  \\
				             & $ \ket{-1}_\text{V}(\frac{1}{2}\sigma_0)_\text{C}$ &  100.5 			&  $98.7\pm .03$   	& 7563   \\
    						 \cline{2-5}	 
           					& gate:  								& 109	                  &  $96.6\pm .01$ & 6339 \\
        \hline
           CROT$_\text{V,N}$      & $\nu=3.13$ MHz&  110       &  $98.0\pm 0.02$	 & 6553 \\	
        \hline
  \end{tabular}
 \caption{  $\pi$-pulse fidelities and times for resonant, square-pulse driving of the 
  		NV+C system at $B=25mT$ for selected states, single- and multi-qubit gates. 
		Gate fidelities where computed from a set of 25, 16 and 8 initial-final state 
		pairs for vacancy, carbon and nitrogen respectively.
		Uncertainties were computed assuming a timing accuracy of $\Delta t=$250ps.
  		}
  \label{tab:medB}
\end{table*}

\begin{figure}
  \includegraphics[width=.45\textwidth]{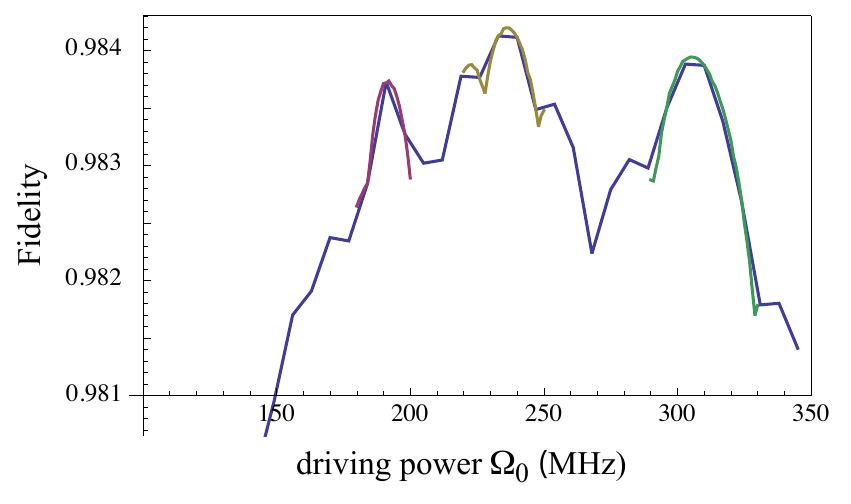} 
   \caption{$\Omega_0$ dependence of the fidelity single-qubit $\pi$-pulse fidelity for driving the vacancy nuclear
   		spin in third-nearest neighbor position. The blue trace is from a rough scan ($\Omega_0$ resolution
		$1$MHz), the other curves were computed with finer resolution to reveal the detailed shape of the maxima.
                  }
     \label{fig:fidvsomega}
\end{figure}

\emph{Driving power. ---}
Our optimization of the driving power $\Omega_0$ yielded complementary results
for nearest and third nearest neighbors. While in the former case, a medium driving power for V and C 
gates and a quite strong power for N yield the highest maximum gate fidelities, the situation is reversed 
for 3rd nearest neighbors.
That third nearest neighbor V driving should be done fast is understandable because we need line-widths 
to be larger than the C-spin level splitting of $C_\parallel\approx 13.5$MHz. This is well satisfied for 
$\Omega_0\geq 150$MHz with sharp peaks in the maximum achievable fidelity occurring whenever the 
phases can be best lined up. 
A resulting $\Omega_0$-dependence of the gate fidelity is Figure~\ref{fig:fidvsomega} for the example of
a (third neighbor) vacancy electronic spin.
While the highest peak in absolute terms occurs at $\Omega_0\approx 240$MHz, taking into account the fidelity 
reduction due to a finite timing accuracy of, assumed, $250$ps shows that the first peak at 192MHz is in fact the 
preferable choice.

However, in an experimental setup, all optimal driving powers identified in this study are rather
technically challenging. Since we consider our system to operate at cryogenic temperatures ($\sim$4-8K), 
sample heating due to the MW and RF-radiation is a serious issue: a rough estimate for the 
maximum permissible 'true' driving power is O($1$W) for which $\pi$-pulse times are roughly 
50ns for the vacancy spin. In our model, this gate time occurs for $\Omega_0=15$MHz, which 
is significantly lower than any of the optimal values we identified (see Table~\ref{tab:medB}). 
This could provide the motivation for an extended search in the low-$\Omega_0$ regime. 
However the difficulty in such a search would be that computation time is proportional to 
gate time and thus roughly inversely proportional to $\Omega_0$. Thus, for all but the vacancy
spin, this would make an extensive search very difficult.

\subsection{Entangling gates}
In the intermediate magnetic field regime, our choice computational basis allows several 
multi-qubit gates to be implemented by a single pulse. The transitions involved are 
indicated schematically in Figure~\ref{fig:medBgates}.

As we see, we can obtain multi-qubit gates between all qubits. 
This set of operations is redundant in that two CNOTs would already be universal, but
this redundancy is very welcome since direct, single-pulse gates are faster and have
a higher fidelity than ones obtained from potentially lengthy gate sequences.  

\begin{figure}
 \hspace{-.4\textwidth} a) \\
 \includegraphics[width=.45\textwidth]{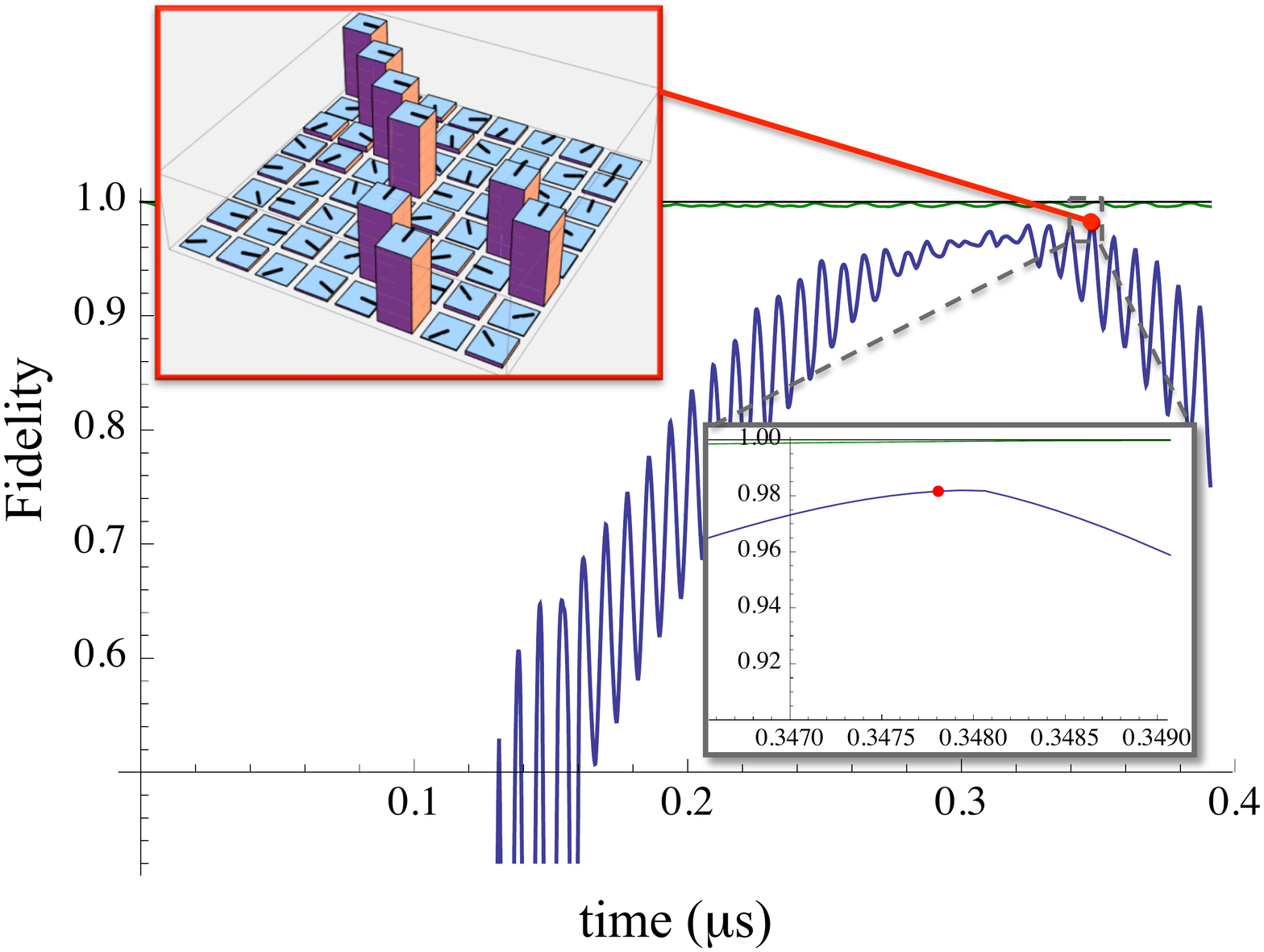} \\
 \hspace{-.45\textwidth} b) \\
 \includegraphics[width=.45\textwidth]{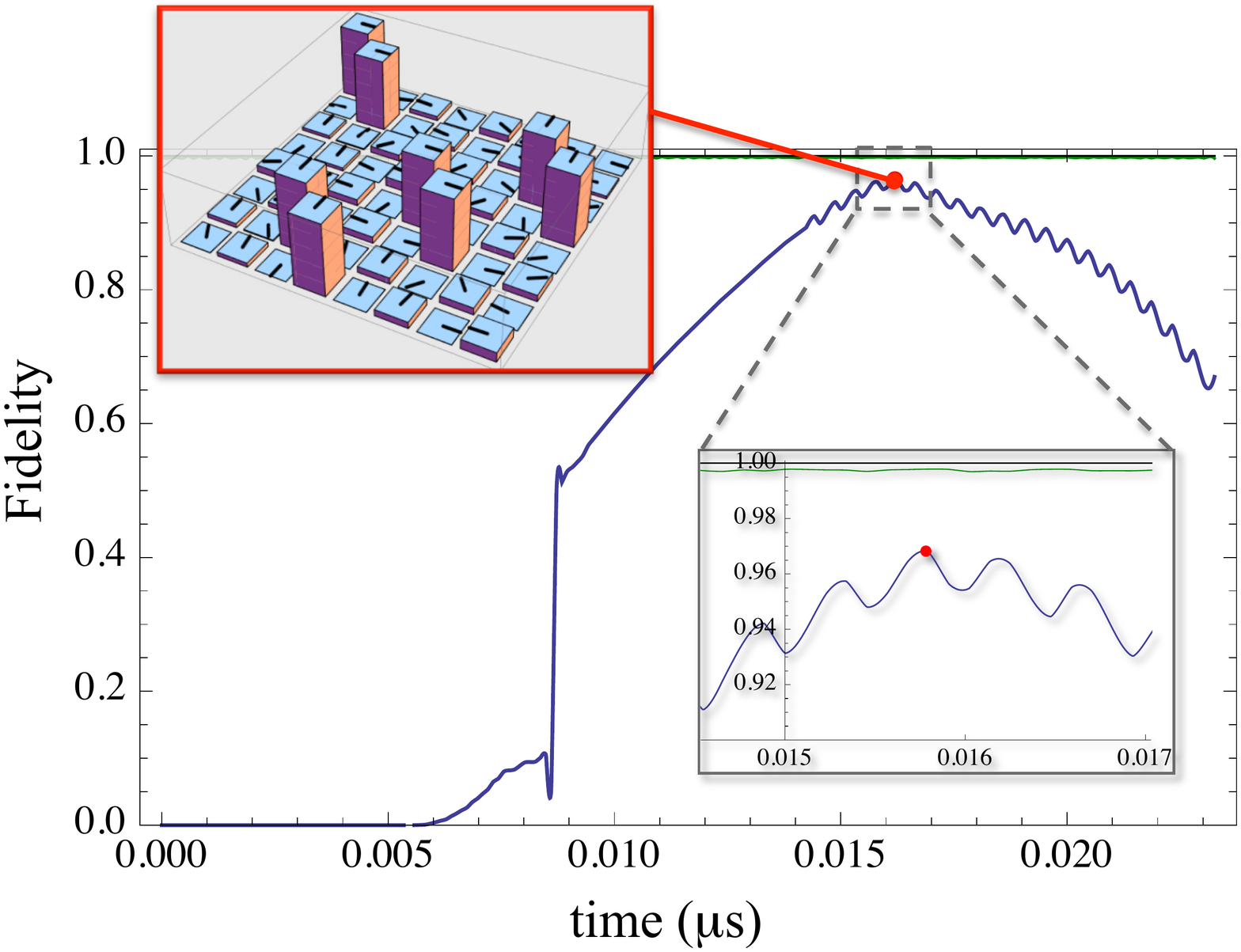}
 \caption{Two examples of two-qubit gates for nearest neighbor carbon: 
 		a) C$_V$ROT$_C$ and b) C$_C$ROT$_V$.
 		The insets in the upper left visualize the unitary transformation
		matrices in the computational basis, where the height of a bar
		stands for the modulus and the small line on top shows the phase.
		(pointing right = phase 0).
                }
   \label{fig:cnots}
\end{figure}

The fidelities we find for the gates along with gate times and optimum driving power $\Omega_0$
are given in Table~\ref{tab:medB} ). Figure~\ref{fig:cnots} shows the
fidelity vs. time and the gate matrices at maximum fidelity for two nearest neighbor 
two-qubit gates.

We should stress, that a gate obtained from 'bare' $\pi$-pulse is not directly
a CNOT but rather a controlled rotation about the axis determined by the phase angle $\phi$ of
the driving field (see the driving Hamiltonian~\eqref{eq:hdrive}). E.g. for an ideal $\pi$-pulse,
the resulting two qubit gate would be $\mathbb{1}_2 \oplus i (\cos(\phi)\sigma_x + \sin(\phi)\sigma_y)$.
For $\phi=0$ this is a CiNOT, necessitating a corrective single-qubit rotation to get an exact CNOT.

In the next section, we will take a closer look at the gates one can derive from this basic
set and see what the expected fidelities are.

\begin{table*}
  \begin{tabular}{l|cc|cc|c}
     gate                                         & \multicolumn{2}{c|}{time ($\mu$s) }           &  \multicolumn{2}{c|}{fid (\%)} & ciruit \\
                                                      &  {\footnotesize n.n.} & {\footnotesize 3rd nb.} & {\footnotesize  n.n.} & {\footnotesize 3rd nb.}    \\
     \hline
     INIT$_\text{V}$                      &  \multicolumn{2}{c|}{100 }           &  \multicolumn{2}{c|}{ 99.9$^{(*)}$ } & \multirow{2}{*}{\includegraphics[scale=.5]{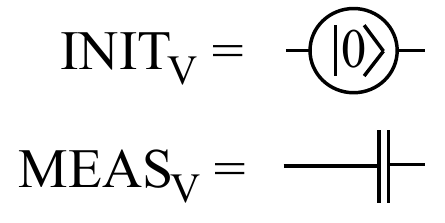} }  \\
     MEAS$_\text{V}$                  &  \multicolumn{2}{c|}{10$\mu$s}        &  \multicolumn{2}{c|}{   "   } &                    \\
                                                      &                                   &       		  &				&			&		\\
     X,Y$_\text{V/C/N}$               &  {\footnotesize.016/.33/6.2}   & {\footnotesize .004/1.0/6.3 } &  {\footnotesize 96.1/98.4/94.1 } & {\footnotesize 97.7/96.9/96.6 } &  \multirow{1}{*}{\includegraphics[scale=.55]{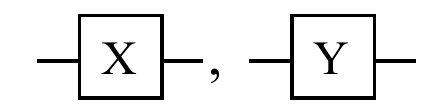}} \\
                                                      &                                   &       		  &				&			&		\\
     CROT$_\text{C,V}$              &     16    &     4ns                &     96.8  &  92.7  &  \multirow{3}{*}{\includegraphics[scale=.5]{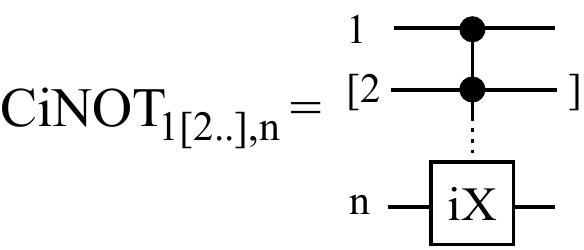}}\\
     CROT$_\text{V,C/N}$          &   0.33/6.23   &  1.08/6.55       & 97.9/91.0   & 98.2/98.0  & \\
     CROT$_\text{CN,V}$          &     0.93   &     0.63            &   94.8  &  97.4  & \\
                                                           &                                   &       		  &				&			&		\\
     \hline
                                                           &                                   &       		  &				&			&		\\
     Z$_\text{V/C/N}$                   &{\footnotesize  .032/.66/12.5}  & {\footnotesize  .008/2.0/12.7}   &  {\footnotesize 92/97/89 } &   {\footnotesize 96/94/93 }   & \multirow{1}{*}{\includegraphics[scale=.55]{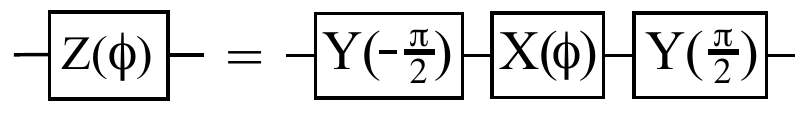}}  \\
                                                           &                                   &       		  &				&			&		\\
     \hline
     $\mathcal{H}_\text{V/C/N}$    & {\footnotesize  .04/.83/15.6 }  &  {\footnotesize .01/2.5/15.9 }    &   {\footnotesize  91 / 96/ 86}&  {\footnotesize 94/ 92 / 92} &  \multirow{1}{*}{\includegraphics[scale=.55]{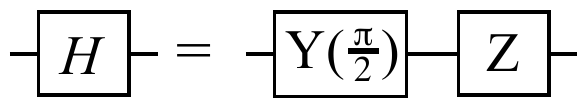}}\\
                                                           &                                   &       		  &				&			&		\\
     CNOT$_\text{C,V}$            &                .35  &  1.06                             &      91  &  94                & \multirow{2}{*}{\includegraphics[scale=.5]{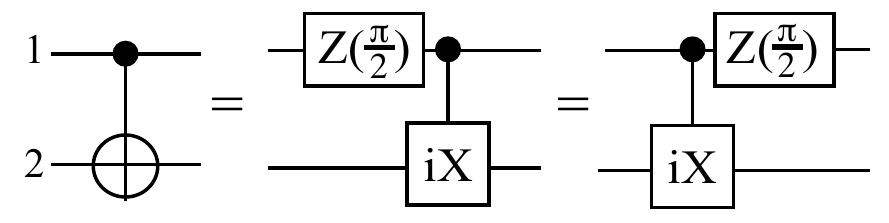}} \\
     CNOT$_\text{V,C}$            &              0.34  &   1.08                        &   90  &  94                  &   \\
     CNOT$_\text{V,N}$            &             12.5  & 12.9                        &      84  &  94                   &  \\ 
                                                    &                         &                              &             &                         &  \multirow{3}{*}{ \includegraphics[scale=.5]{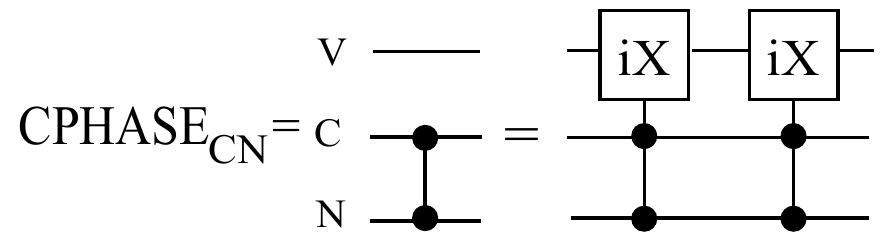}}   \\
     CPHASE$_\text{C,N}$      &               1.86   &  1.26                 &      90  &  95                   &  \\
                                                     &                         &                              &             &                         &  \\
                                                    &                         &                              &             &                         &  \\
     \hline
     CNOT$_\text{N,V}$            &            43.7        & 44.6              &        51   &  70        & \\   
                                                    &                           &                        &             &                         &  \multirow{2}{*}{\includegraphics[scale=.55]{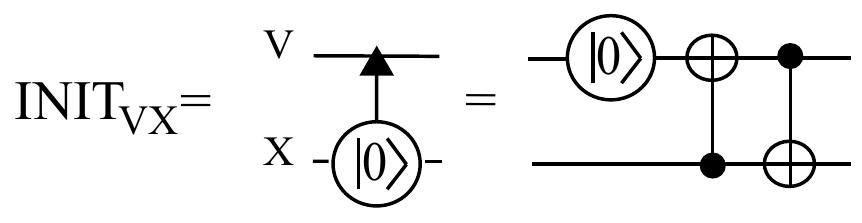}}   \\
     INIT$_\text{C}$                    &           .79          &   2.24                &        85   &  82        &    \\   
     INIT$_\text{N}$                    &           56.3        &    57.6            &   $<50$ & 66         &    \\   
                                                    &                         &                              &             &                         &  \multirow{2}{*}{\includegraphics[scale=.55]{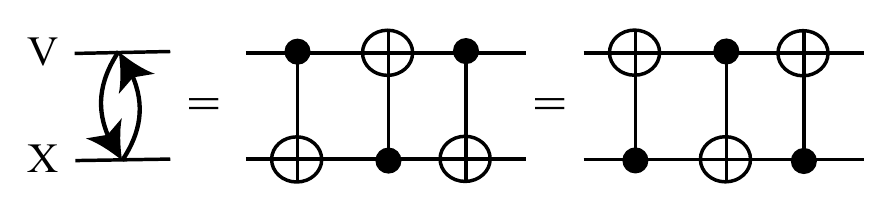}}    \\
     SWAP$_\text{VC}$             &           1.03        &   3.20             &       79   &  71     &   \\   
     SWAP$_\text{VN}$             &          68.7         &     70.4              &  $<$50 &  61        &  \\   
                                                    &                         &                              &             &                 &  \multirow{2}{*}{\includegraphics[scale=.55]{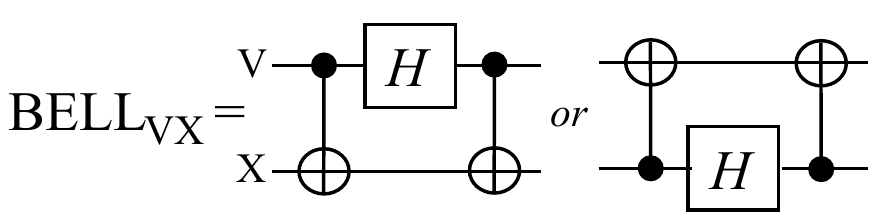}} \\
     BELL$_\text{VC}$              &             1.5         &   4.6              &         84 &  70        &    \\   
     BELL$_\text{VN}$              &            25.0       & 25.8              &         64  &  83       &  \\  
                                                   &                         &                              &             &                         &  \\
     \hline
     BELLM$_\text{VC}$          &        2.75           &  8.0              &           67    &  50     &  \multirow{2}{*}{\includegraphics[scale=.55]{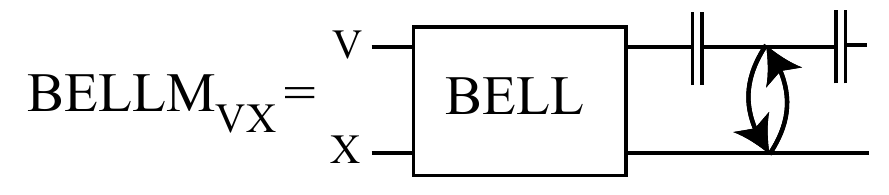}}   \\   
     BELLM$_\text{VN}$          &       93 .9           & 96.4                &    $<50$ & 50  & \\  
                                                    &                         &                              &             &                         &  \\
  \end{tabular}
  \caption{A list of the relevant gates in the NV+C system with gate times and expected fidelities.
                  Values for primitive gates are taken from the results in Section 4, which in turn are
                  used to estimated those of derived gates.      
                  (*) : assumed values
                }
   \label{tab:dgatefids}
\end{table*}

\section{Derived gates and sequences}
\label{sec:dgates}
In the previous section we looked at gates and operations implementable with a single pulse. 
Here we want to extend this to sequences of pulses in order to realize a set of useful gate operations
on the three qubit NVC system at intermediate magnetic field. 
In Figure~\ref{fig:dgates}, we show an overview of relevant gates in the NV system both primitive and
derived ones together with the dependency structure.

The primitives presented in the previous section include all single-qubit rotations about an axis in
the x-y plane (from which one can construct z-rotations), as well as the four entangling operations 
CROT$_\text{V,C}$, CROT$_\text{C,V}$, CROT$_\text{VN}$ and CROT$_\text{CN,V}$,
where CROT$_\text{1[2],3}$ denotes a conditional rotation applied to qubit 3 controlled by the state
of qubit(s) 1 (and 2). For instance, if one chooses to perform an X($\pi/2$) rotation, \ie a $\pi$-pulse 
about the x-axis, the resulting operation would be a CiNOT, which is equivalent to a CNOT up to a 
$\pi/2$ z-rotation on the control qubit.
In addition, one has the non-unitary initialization of the vacancy spin into the $\ket{0}_\text{V}$ state. 
The standard technique at room temperature is to employ off-resonant excitation with green 
laser light, and was used in virtually all NV experiments to date. However, at low temperature 
resonant driving to a state with preferential decay to the $m_S=0$ state, \eg $\ket{A_2}$ is much 
faster and one should be able to reach high fidelities after only a few cycles.  

These primitives clearly form a universal set which has in fact some redundancy. For instance, 
we need only one out of CROT$_\text{V,C}$ and CROT$_\text{C,V}$ as well as CROT$_\text{VN}$ and CROT$_\text{CN,V}$.
Having them all at our disposal potentially improves both gate time and fidelity. 
Table~\ref{tab:dgatefids} gives an overview of time and fidelity for the gates shown in 
Figure~\ref{fig:dgates}. It is clear that all gates involving the nitrogen nuclear spin are both slow 
and low fidelity, so unless this can be resolved by further optimization of square pulses or more 
advanced pulse shaping, it is best to try and work without it.
Excluding the degree of freedom of the nitrogen means we are reduced to a two-qubit system. 
Thus it is no longer possible to perform any error correction within the device unless we introduce
another $^{13}$C. However, in such a case we expect similar problems as with the nitrogen.
Thus, its use in, \eg repeaters would depend on the initial entangling link being high-fidelity in 
the first place. Ways to establish such links probabilistically have been proposed~\cite{Childress-05pra052330,Nemoto-13arx1309.4277} 
using state dependent reflectivity of cavities together with path-erasure techniques. 

If we assume entanglement links between two NVC systems are established with fidelity 
exceeding 99.9\% using this method, a Bell measurement could be performed with fidelity 
$f_\text{BELL}=$70\% (74\%) (cp. Table~\ref{tab:dgatefids}) allowing only a single
round of entanglement swapping before link fidelity drops below the classical threshold. 
This shows that for strongly coupling carbon $^{13}$C it is necessary to go beyond the
square pulse paradigm and consider shaped pulses and pulse sequences. In technical 
applications, this would mean a complication that is avoidable in bare NV centers, where 
square pulses are already good enough. However, we think it is still interesting to pursue 
this course, as the carbon offers a single-qubit gate speed-up by a factor of more than 10 
for nearest- and still about 5 for third-nearest-neighbors.

\section{Conclusion}
We numerically investigated a system consisting of an $^{15}$NV$^-$ center and a nearby,
strongly hyperfine-coupled carbon $^{13}$C nuclear spin in two different magnetic field regimes.
Within a conservative yet realistic model, we determined the achievable fidelities for specific 
states as well as gates using only simulated square pulses of microwave and radio-frequency 
radiation.  
We find that in the low magnetic field regime only some special starting and target state 
combinations allow high fidelity operations. This suggests that careful selection of states 
gives us sufficient fidelity to perform some quantum information tasks.
Gate fidelity suffers from the limited state z-fidelity and level separation. 
The situation is much better at intermediate fields. There, we found fidelities of up to 98\% 
for single-qubit gates on the carbon nuclear spin and 97\% for the
vacancy electronic spin. The nitrogen single-qubit as well as multi-qubit gate fidelities are
somewhat lower than that. 
If we analyze the expected gate times of gates derived from these primitives via
straightforward concatenation, we find that using a strongly bound carbon does indeed 
offer potential speed up of operations. However the fidelities of these derived gates quickly 
deteriorateswith nesting level. 
Thus this study indicates that gates implemented via square pulses can be used only in 
limited applications. For general applications going beyond the square pulse 
paradigm and using pulse-shaping techniques like optimal control is required.

\section{Acknowledgement} 
We thank M.S. Everitt, S.J. Devitt and H. Kosaka for valuable comments and discussions. 
This research was partially supported under the Commissioned Research of National 
Institute of Information and Communications Technology (NICT) (A \& B) project.
   
%\bibliographystyle{apsrev}
%\bibliography{../../../misc/bib/nvc,../../../misc/bib/qcnqi}

\appendix
\section{Measuring fidelity}
\label{app:fidmeas}
The fidelity between two quantum states described by density matrices
$\rho$ and $\sigma$ can be determined by
\begin{equation}
  F(\rho,\sigma) = \left(\Tr\sqrt{\sqrt{\rho}\sigma\sqrt{\rho}}\right)^2 .
\end{equation}
The square-root is defined for hermitian operators and can be computed
from the eigenspectrum via $\sqrt{\rho} = \text{diag}\{\sqrt{\rho_1}, ...,\sqrt{ \rho_D}\}$
where $D=\text{dim}\,\mathcal{H}$ and $\rho_n$ are the eigenvalues
of $\rho$.
If one of the states, say $\sigma$, is a pure state,
this simplifies to
\begin{equation}
  F(\rho, \sigma = \ket{\phi}\!\bra{\phi}) = \Tr\left(\rho\ket{\phi}\!\bra{\phi} \right)= \braket{\phi | \rho | \phi}\, .
\end{equation}

For purely unitary time evolutions, the exact gate fidelity can be computed by just considering 
(the time evolution of) an ONB of the Hilbert space $\mathcal{H}$. In practice, this would 
require computing the time evolution for $D$ different starting states.
However, since $\mathscr{E}_t[\rho]$, the actual time evolution of the system, is dissipative in 
our case, the exhaustive description necessary for calculating the exact 
gate fidelity requires computing the time evolution for all $D^2$ generators 
of the space of hermitian operators on $\mathcal{H}$. 
As a standard way of assessing gate-fidelity this is computationally 
too costly even for our modest Hilbert-space size of $D=12$. 

Therefore, we settled on the practical solution of computing the state fidelities 
for a suitably large subset of states from $\mathcal{H}$ and taking as our gate 
fidelity the minimum among all the obtained values. The size of these sets were 25, 16 and 8 states 
when assessing vacancy, carbon and nitrogen driving pulses respectively.
In detail, the state sets were
\begin{align*}
  \text{Vacancy (25 states): } & \left\{ \ket{0, k , l}, \ket{0}\otimes(\rho_\text{mixed})_\text{C} \otimes (\rho_\text{mixed})_\text{N} \right\} \\
    \text{Carbon (16 states): } & \left\{ \ket{m, \dn, l} \right\} \\
     \text{Nitrogen (8 states): } & \left\{ \ket{ n, l, \dn} \right\} \; ,
\end{align*}
where $k \in \left\{ \up, \dn, x\pm, y\pm \right\}$, $l \in \left\{ \up, \dn, x\pm \right\}$, $m \in \left\{ 0, -1, x\pm \right\}$
and $n \in \left\{ 0, -1 \right\}$.

\section{Numerical simulation} 
\label{app:model}
\emph{Decoherence model. ---} We implemented a master equation in Lindblad form
\begin{equation}
  \dot{\rho} = - \frac{i}{\hbar} \left[ H_\text{VC}, \rho\right] + \sum_{k}\,\gamma_k \left( L_k\rho L_k^\dagger -\frac{1}{2} \left\{ L_k^\dagger L_k, \rho \right\}  \right) .
  \label{eq:meq}
\end{equation}
Here $k$ runs over all subsystems and diagonal/off-diagonal elements, \eg $k=(\text{C},1+)$ 
labels the raising operator for the carbon spin, which together with $(\text{C},1-)$ is responsible 
for carbon spin relaxation.
Thus, the Lindblad operators $L_k$ describe relaxation and dephasing of each system (V, C and N) 
individually and the $\gamma_k$ are the inverses of the experimentally observed relaxation and 
decoherence times $T_1$ and $T_2$ of the individual subsystems except for vacancy dephasing 
rates $\gamma_{2,\text{V} a/b}$. This we chose time dependent, to reproduce the experimentally 
observed Gaussian (and thus non-Markovian) dephasing of the vacancy electron spin. 
This kind of dephasing is observed for the time evolution of a spin which evolves under the influence 
of a weak and randomly varying magnetic field, which in case of the NV stems from other spins in 
the vicinity (the electronic spins of nitrogen P1 centers as well as carbon $^{13}$C nuclear spins).
Gaussian dephasing is obtained for a linear time dependence of $\gamma_{2\text{V} a/b}= t/T_{2,\text{V}}^2$.

\emph{Simulation. ---} Numerical simulations where performed in \emph{Mathematica} (version 7.0) using the 
built-in NDSOLVE function to integrate the Master equation~\eqref{eq:meq} up to the desired final
time starting in some state $\rho(0)=\rho_0$ of the entire system. Single qubit gate fidelities where
computed as described in the previous section while multi-qubit gates where extracted in a similar
fashion, however comparing each final state to all other target states in addition to the desired one.

\section{Derived gates}
\label{app:dgates}
\begin{figure}
 \includegraphics[scale=.37]{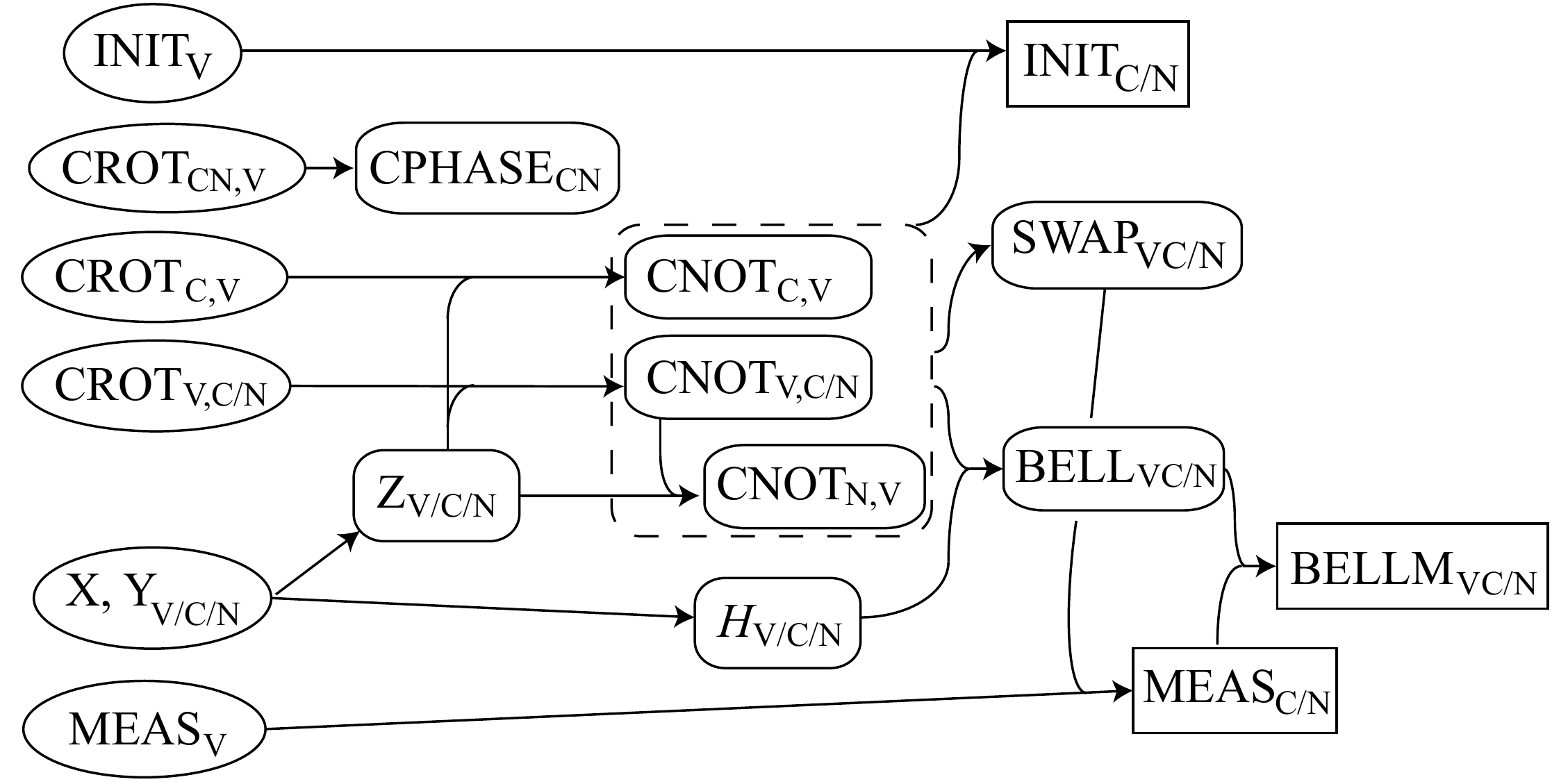} \\
  \caption{Graph illustrating the dependencies of derived gates on primitive ones (left side).
                 }
   \label{fig:dgates}
\end{figure}

 In the intermediate field regime ($B=25$mT) we computed the fidelities of some interesting derived 
 gates based on the simulation results obtained for primitive gates. 
 Derived gates are constructed from sequences of primitive ones according to some gate identity. 
 Following the prescription of these identities we obtain derived gate parameters by multiplying the 
 fidelities and summing the times of the constituent primitives. This is consistent with the limitation 
 of the NV$^-$+C system where gates cannot be performed in parallel on different subsystems for 
 physical reasons, even though this might be possible logically (e.g. single-qubit gates on different 
 qubits commute).
 For dependency between primitives and derived gates, see Figure~\ref{fig:dgates}, for the complete
 list of gates and the (highest fidelity) identities see Table~\ref{tab:dgatefids}.

 Frequently there are several different ways to obtain a given gate, in particular since our set of 
 primitives is redundant. For instance one can obtain a CNOT$_\text{C,N}$  either by applying the 
 square of a CNOT$_\text{CN,V}$ (=TOFFOLI$_\text{CN,V}$) sandwiched between two Hadamard 
 gates on the nitrogen or, alternatively, via a CNOT$_\text{V,N}$ sandwiched between two 
 SWAP$_\text{VC}$. 
 In this case the former is clearly the faster and higher fidelity alternative. 
 However there are also cases where one has to choose between fidelity or speed. For example 
 a BELL$_\text{VC}$ gate can be achieved either via CNOT$_\text{V,C}\cdot$H$_\text{V}\cdot$CNOT$_\text{V,C}$
 or with the same but with V and C switching roles. We must point out that these two options do
 in fact not give the exact same gate: the former realizes the basis-state mapping 
   $\ket{00}\rightarrow\ket{\psi_+}$, $\ket{01}\rightarrow\ket{\phi_+}$, $\ket{10}\rightarrow\ket{\phi_-}$, $\ket{11}\rightarrow\ket{\psi_-}$
 while the latter has instead  $\ket{01}\rightarrow\ket{\phi_-}$, $\ket{10}\rightarrow\ket{\phi_+}$,
 where $\ket{\phi_\pm}$ and $\ket{\psi_\pm}$ denote the even and odd parity Bell states respectively.
 But both map the computational basis onto a Bell basis, and the permutation between the Bell vectors
 just requires a slightly different interpretation of measurement results and we can thus regard
 them as effectively equivalent.
 But in practice it makes a great difference which one we choose to perform: the former gate identity involves
 two slow, but higher fidelity CNOTs and one fast Hadamard and vice versa for the latter. 
 Gate times are (for nearest neighbor) $860$ns versus $720$ns while the fidelities 
 are 84\% compared to 74\%.

\end{document}